\documentclass[aps,pre,twocolumn,superscriptaddress,10pt]{revtex4-1}
\usepackage{graphicx}
\usepackage{subfigure}
\usepackage{amsmath}
\usepackage{amsfonts}
\usepackage{amssymb}
\usepackage{amsthm}
\usepackage{times}
\usepackage[colorlinks,citecolor=blue,linkcolor=blue]{hyperref}
\usepackage{color}
\usepackage{setspace} 
\usepackage{verbatim}
\newcommand{\bea}{\begin{eqnarray}}
\newcommand{\eea}{\end{eqnarray}}

\begin{document}
\title{Thermodynamic uncertainty relation in quantum thermoelectric junctions}  

\newcommand{\UT}{\affiliation{Department of Chemistry and Centre for 
Quantum Information and Quantum Control, 
University of Toronto, Toronto, Ontario, M5S 3H6, Canada}}

\author{Junjie Liu}
\UT
\author{Dvira Segal}
\UT

\begin{abstract} 
Recently, a thermodynamic uncertainty relation (TUR) has been formulated for classical
Markovian systems demonstrating trade-off between precision (current fluctuation) 
and cost (dissipation). Systems that violate the TUR are interesting as they 
overcome another trade-off relation concerning the efficiency of a heat engine, 
its power, and its stability (power fluctuations).
Here, we analyze the root, extent, and impact on performance 
of TUR violations in quantum thermoelectric junctions at steady state.
Considering noninteracting electrons, first we show that only 
the ``classical" component of the current noise, arising from single-electron 
transfer events follows the TUR. The remaining, ``quantum" part of current noise 
is therefore responsible for the potential violation of TUR in such quantum systems. 
Next, focusing on the resonant transport regime we
determine the parameter range in which the violation of the TUR can be observed---for both voltage-biased junctions and thermoelectric engines.
We illustrate our findings with exact numerical simulations of a serial double quantum dot system.
Most significantly, we demonstrate that the TUR always holds in noninteracting thermoelectric generators when approaching the thermodynamic efficiency limit.
\end{abstract}

\date{\today}

\maketitle
\section{Introduction}

Fluctuations are ubiquitous in small systems away from equilibrium. 
Identifying universality in the behavior of fluctuations is thus 
central to the development of modern nonequilibrium thermodynamics and statistical mechanics. 
Recently, a remarkably simple cost-precision trade-off relation, 
coined the ``thermodynamic uncertainty relation" (TUR) 
had been formulated for classical Markovian systems in non-equilibrium steady state
\cite{Barato.15.PRL,Gingrich.16.PRL,Pietzonka.16.PRE,Polettini.16.PRE,Pietzonka.16.JPA,Gingrich.17.JPA,Seifert.18.PA},
\bea
\frac{ \langle \langle j_{\alpha}^2\rangle \rangle} {\langle j_{\alpha}\rangle^2} \sigma \geq 2.
\label{eq:tur}
\eea
Here, $\langle j_{\alpha} \rangle$ is the averaged current of e.g. particle number or energy and
$\langle\langle  j_{\alpha}^2\rangle\rangle=$ $\langle  j_{\alpha}^2\rangle-\langle  j_{\alpha}\rangle^2$
 corresponds to the second cumulant of the current. $\sigma$ is the average entropy production rate, $k_B=1$.

Manifesting as an inequality, the TUR [Eq. (\ref{eq:tur})] implies that 
a more precise output requires higher entropy production (cost). 
Given its fundamental and conceptual importance, the TUR has 
been refined \cite{Macieszczak.18.PRL,Hasegawa.19.A} and generalized to finite times \cite{Pietzonka.17.PRE,Horowitz.17.PRE,Manikandan.18.JPA}, discrete time and periodic dynamics \cite{Proesmans.17.EPL,Chiuchiu.18.PRE,Barato.18.NJP,Koyuk.18.JPA}, multidimensional systems \cite{Dechant.19.JPA}, and bounds on counting observables and first-passage times \cite{Garrahan.17.PRE,Gingrich.17.PRL}, with applications to biochemical motors \cite{Hwang.18.JPCL}, heat engines \cite{Pietzonka.18.PRL,Holubec.18.PRL,Dechant.18.JSM,Dechant.19.JPA} and a variety of nonequilibrium problems \cite{Hyeon.17.PRE,Pigolotti.17.PRL,Brandner.18.PRL}.
Specifically, for an engine operating in a nonequilibrium steady state, the TUR translates into a
trade-off relation between the output power, power fluctuations, and the engine's efficiency: 
According to the bound, power fluctuations diverge when operating an engine at 
finite power while approaching the Carnot efficiency \cite{Pietzonka.18.PRL}.

In the past year, there has been a great deal of interests on applying the TUR 
to quantum systems, or alternatively, in formulating a new quantum mechanical 
bound \cite{Guarnieri.19.A}.
In particular, it has been demonstrated that the TUR can be violated in the quantum regime 
\cite{Bijay.18.PRB,Ptaszynski.18.PRB} in non-Markovian problems.

This finding immediately opens up several interesting perspectives. 
On the one hand, one can anticipate the reduction of fluctuations in certain quantum systems and 
hence harness quantum effects to circumvent the trade-off relations on power and efficiency of 
thermodynamic engines \cite{Shiraishi.16.PRL,Pietzonka.18.PRL,Holubec.18.PRL}, 
thereby enhancing the performance of quantum engines. 
On the other hand, the violation of the TUR suggests on the existence of intrinsic 
quantum bounds on precision. Notably, a recent study showed that the geometry of 
quantum steady states implied a looser bound on precision \cite{Guarnieri.19.A}. 
Despite of this progress, the applicability of the ``classical" TUR, Eq. (\ref{eq:tur}) in the 
quantum regime still remains largely unexplored.
Specifically, mechanisms responsible for the violation of the TUR are still not fully understood 
even in simple quantum systems. Moreover, in quantum engines with multiple thermodynamic affinities, 
one may expect large fluctuations and thus the validity of the TUR.

In this work, we focus on noninteracting quantum thermoelectric junctions at steady state.
Our objectives are (i) to uncover the origin of TUR violations in such quantum transport models, 
(ii) to identify the range of parameters where violation can take place, and (iii) 
to assess the impact of TUR violation on the performance (power-fluctuations-efficiency) 
of thermoelectric generators.
Our analysis is based on the exact full counting statistics of currents 
\cite{Levitov.93.JETP,Levitov.96.JMP}, which allows us to explore the thermodynamic quantities 
involved in the TUR in an exact manner, without compromising the validity regime of our conclusions.  

Our work resolves several issues. First, by splitting the current noise into two 
kinds,``classical" noise that results from single electron transfer events, and ``quantum" noise, which 
reflects correlated exchange processes of two electrons, we show that only the ``classical" noise 
definitely satisfies the TUR. Thus, the violation of the TUR in our modeling can be solely 
attributed to the existence of ``quantum" noise. 

Second, we focus on the resonant tunnelling regime, where analytic expressions are available. 
Here, we determine the voltage range within which the violation of the TUR can be observed in charge transport
systems. We show that this voltage range can be modified by adjusting the 
chemical potentials of the metal leads relative to the Fermi energy, 
thereby offering a promising mean to enhance or suppress the ``quantum" noise at will. 
In systems operating as thermoelectric engines, we demonstrate that the violation of the TUR 
only occurs within a narrow parameter regime constrained by the temperature of the 
hot bath as well as the energies of the resonance (realized with quantum dots or molecular orbitals).

We illustrate our findings using the serial double quantum dot (DQD) system. 
In charge transport setups, exact simulations confirm our theoretical expectation---that one can 
observe the violation of the TUR within a specific voltage range, 
which is sensitive to the partitioning of the chemical potentials. 
In systems working as thermoelectric engines, we confirm from simulations 
that violations occur in the resonant tunnelling regime within a narrow range of parameters. 
Even so, the TUR is restored when the efficiency of the thermoelectric engine 
approaches its thermodynamic Carnot limit. 
Beyond that, violations can only occur when the system no longer behaves as a thermoelectric generator.

Our central conclusion is that in the resonant tunnelling regime, 
noninteracting thermoelectric engines  {\it can violate} the trade-off relation between efficiency, 
power and constancy \cite{Pietzonka.18.PRL}, but quantum effects cannot be utilized to enhance 
the performance {\it when approaching the thermodynamic efficiency limit} as the engines always respect the TUR in that limit, in agreement with a recent study \cite{Samuelsson}. 
Note that in Ref. \cite{Ptaszynski.18.PRB}, violation of TUR were demonstrated in a 
thermoelectric junction, but in a regime where the system does not produce power.

The paper is organized as follows. 
We describe the noninteracting thermoelectric junction model and the TUR in Section \ref{sec:2}. 
In Section \ref{sec:3}, we derive conditions under which the violation of the TUR can be observed.
In Section \ref{sec:4}, we illustrate our findings using the serial DQD system, and present numerical 
results. We summarize our findings in section \ref{sec:5}.


\section{Noninteracting thermoelectric junctions}\label{sec:2}

\subsection{Current and noise}

We consider quantum thermoelectric junctions with a multilevel, noninteracting 
system sandwiched between two metal leads characterized by different chemical potentials 
and temperatures. The energy and charge transport characteristics of thermoelectric junctions 
are fully described by their joint energy and particle full counting statistics. 
In particular,  
if each metal is coupled to the system through a single molecular orbital, 
the steady-state cumulant generating function (CGF) associated with the charge (C) 
and energy (E) currents can be exactly formulated, given 
by a generalized Levitov-Lesovik formula 
\cite{Levitov.93.JETP,Levitov.96.JMP,Esposito.15.PRBa} (setting $k_B=1$, $\hbar=1$)
\begin{eqnarray}
\label{eq:CGF}
&&G(\{\chi\}) = \int_{-\infty}^{\infty}\frac{d\epsilon}{2\pi}\ln\Big(1+\mathcal{T}(\epsilon)\{f_L(\epsilon)[1-f_R(\epsilon)] 
\nonumber\\
&&\times
[e^{i(\chi_C+\epsilon\chi_E)}-1]
\left.+f_R(\epsilon)[1-f_L(\epsilon)][e^{-i(\chi_C+\epsilon\chi_E)}-1]\}\right).
\nonumber\\
\end{eqnarray} 
Here, $\{\chi\}=\{\chi_C,\chi_E\}$ are counting fields for charge and energy transfer processes. 
$\mathcal{T}(\epsilon)$ is the energy-dependent transmission coefficient determined 
by the retarded and advanced Green's functions of the system in the absence of counting fields. 
$f_v(\epsilon)=[e^{\beta_v(\epsilon-\mu_v)}+1]^{-1}$ is the Fermi distribution function 
for the two metal leads $v=L,R$ with chemical potential $\mu_v$ and inverse temperature $\beta_v$.

The charge and energy mean currents, and their corresponding current noises can be directly obtained 
from the above CGF as 
$\langle j_{\alpha}\rangle=\left.\frac{\partial G}{\partial(i\chi_{\alpha})}\right|_{\{\chi\}=0}$ 
and 
$\langle\langle j^2_{\alpha}\rangle\rangle=\left.\frac{\partial^2G}{\partial(i\chi_{\alpha})^2}\right|_{\{\chi\}=0}$, respectively, where $\alpha=C, E$.  Explicitly, the mean currents read
\begin{equation}
\label{eq:jj}
\langle j_{\alpha}\rangle~=~\int_{-\infty}^{\infty}\frac{d\epsilon}{2\pi}\xi_{\alpha}\mathcal{T}(\epsilon)[f_L(\epsilon)-f_R(\epsilon)],
\end{equation}
where $\xi_{\alpha}=1$ $(\epsilon)$ for $\alpha=C$ $(E)$. 
By convention, the sign of the current is taken positive if it flows from the left to the right lead.
The noises read
\begin{eqnarray}
\label{eq:noise}
\langle\langle j^2_{\alpha}\rangle\rangle &=& \int_{-\infty}^{\infty}
\frac{d\epsilon}{2\pi}\xi_{\alpha}^2
\Big(\mathcal{T}(\epsilon)\left\{f_L(\epsilon)\left[1-f_L(\epsilon)\right]\right.
\nonumber\\
&&\left.+f_R(\epsilon)\left[1-f_R(\epsilon)\right]\right\}\nonumber\\
&&+\mathcal{T}(\epsilon)\left[1-\mathcal{T}(\epsilon)\right]\left[f_L(\epsilon)-f_R(\epsilon)\right]^2\Big).
\end{eqnarray}
Below, we refer to ``charge transport junctions" as steady state 
setups with $\beta_L=\beta_R$ but $\mu_L\neq \mu_R$.
``Thermoelectric junctions" are driven by two thermodynamics forces, 
with $\beta_L\neq\beta_R$ and $\mu_L\neq \mu_R$;  in 
``thermoelectric engines" or ``generators", power is produced.

\subsection{Thermodynamic Uncertainty Relation}

We introduce the relative uncertainty of each individual current as
\begin{equation}
\Phi_{\alpha}~\equiv~\frac{\langle\langle j_{\alpha}^2\rangle\rangle}{\langle j_{\alpha}\rangle^2}.
\end{equation}
%
Quite remarkably, the relative uncertainty together with the mean entropy production rate
$\sigma$ obey the so-called thermodynamic uncertainty relation 
in classical Markovian systems \cite{Barato.15.PRL,Gingrich.16.PRL},
\begin{equation}\label{eq:tur_classical}
\sigma\Phi_{\alpha}\ge 2.
\end{equation}
Namely, the product $\sigma\Phi_{\alpha}$ is bounded from below by 2.

To explore the possible violation of the TUR in  nanojunctions, 
we analyze the full expression for the current noise
(\ref{eq:noise}) and partition it into two
terms, which  we loosely refer to as ``quantum" (qu) and ``classical" (cl) noise. 
Below we show that the classical part of the noise obeys the TUR, 
thus only the quantum part can be responsible for TUR violations.

The current noise 
Eq. (\ref{eq:noise}) can be divided into two components, 
$\langle\langle j^2_{\alpha}\rangle\rangle=\langle\langle j^2_{\alpha}\rangle\rangle_{cl}-\langle\langle j^2_{\alpha}\rangle\rangle_{qu}$, with 
\cite{Nazarov.09.NULL} 
\begin{eqnarray}
\langle\langle j^2_{\alpha}\rangle\rangle_{cl} &=&  \int_{-\infty}^{\infty}\frac{d\epsilon}{2\pi}\xi_{\alpha}^2\mathcal{T}(\epsilon)\{f_L(\epsilon)[1-f_R(\epsilon)]\nonumber\\
&&+f_R(\epsilon)[1-f_L(\epsilon)]\},\nonumber\\
\langle\langle j^2_{\alpha}\rangle\rangle_{qu} &=& \int_{-\infty}^{\infty}\frac{d\epsilon}{2\pi}\xi_{\alpha}^2\mathcal{T}^2(\epsilon)[f_L(\epsilon)-f_R(\epsilon)]^2.
\end{eqnarray}
The ``classical" term $\langle\langle j_{\alpha}^2\rangle\rangle_{cl}$
depends on the single-electron transmission function. Thus, it is regarded as the quantum 
analogue of the classical expression to the noise---with additional factors accounting for the exclusion 
principle. In contrast, the ``quantum" contribution $\langle\langle j_{\alpha}^2\rangle\rangle_{qu}$ 
has no classical counterpart as it is second order in the transmission coefficient, 
and thus describes the correlated exchange of two electrons.  
Using this decomposition, the relative uncertainty is organized as follows,
\begin{equation}
\Phi_{\alpha}~=~\Phi_{\alpha}^{cl}-\Phi_{\alpha}^{qu},
\end{equation}
where $\Phi_{\alpha}^{cl}=\langle\langle j_{\alpha}^2\rangle\rangle_{cl}/\langle j_{\alpha}\rangle^2$ and $\Phi_{\alpha}^{qu}=\langle\langle j_{\alpha}^2\rangle\rangle_{qu}/\langle j_{\alpha}\rangle^2$.

We now prove that the classical noise satisfies the TUR.
Since we only consider systems with time-reversal symmetry, 
we introduce the following quadratic functional for the classical noise \cite{Brandner.18.PRL},
\begin{equation}\label{eq:theta}
\Theta_{\alpha}(x)~=~\sigma+4\left(\langle j_{\alpha}\rangle x+\langle\langle j^2_{\alpha}\rangle\rangle_{cl} x^2/2\right),
\end{equation}
where $x$ is a real parameter. We recall that the entropy production is written as
\begin{equation}\label{eq:sigma_ge}
\sigma~=~\sum_{\alpha=C,E}\mathcal{F}_{\alpha}\langle j_{\alpha}\rangle,
\end{equation}
with thermodynamic affinities 
$\mathcal{F}_E=\beta_R-\beta_L$ and $\mathcal{F}_C=\beta_L\mu_L-\beta_R\mu_R$. 
We define
 $\mathcal{D}\equiv\xi_{C}\mathcal{F}_{C}+\xi_{E}\mathcal{F}_{E}$;
recall that $\xi_C=1$ and $\xi_E=\epsilon$. We  note that
\bea
&&f_L(\epsilon)[1-f_R(\epsilon)]+f_R(\epsilon)[1-f_L(\epsilon)]
\nonumber\\
&&=f_R(\epsilon)[1-f_L(\epsilon)](e^{\mathcal{D}}+1),
\nonumber\\
&&f_L(\epsilon)-f_R(\epsilon)=f_R(\epsilon)[1-f_L(\epsilon)](e^{\mathcal{D}}-1).
\eea
We can therefore express the quadratic functional $\Theta_{\alpha}(x)$ as
\begin{eqnarray}
\Theta_{\alpha}(x) &=& \int_{-\infty}^{\infty}\frac{d\epsilon}{2\pi}\mathcal{T}(\epsilon)f_R(\epsilon)[1-f_L(\epsilon)]
\nonumber\\
&\times&
\left\{\mathcal{D}(e^{\mathcal{D}}-1)\right.
\left.+4 x\xi_{\alpha}(e^{\mathcal{D}}-1)+2x^2\xi_{\alpha}^2(e^{\mathcal{D}}+1)\right\}.
\nonumber\\
\end{eqnarray}
Minimizing the term inside the curly bracket with respect to $x$ yields 
$\mathcal{D}(e^{\mathcal{D}}-1)-2(e^{\mathcal{D}}-1)^2/(e^{\mathcal{D}}+1)$, 
which is non-negative by noting that $y(e^y-1)\ge2(e^{y}-1)^2/(e^y+1)$ for any real $y$. 
Hence, the quadratic functional $\Theta_{\alpha}$ is positive semidefinite for 
any $x$, since $\mathcal{T}(\epsilon)$ and $f_R(\epsilon)[1-f_L(\epsilon)]$ are non-negative. 

Back to the original form, Eq. (\ref{eq:theta}), taking the minimum with respect to $x$ yields
\begin{equation}
\label{eq:tur_cl}
\sigma\Phi_{\alpha}^{cl}\ge 2.
\end{equation}
Altogether, by extending the analysis of Ref. \cite{Brandner.18.PRL} 
to systems with multiple thermodynamic affinities,  
we rigorously show that for time-reversible quantum thermoelectric junctions the 
classical component of the relative uncertainty $\Phi^{cl}_{\alpha}$ always satisfies the TUR. 
As for the quantum component $\Phi^{qu}_{\alpha}$, although it is in general nonzero in quantum systems, 
we conclude that the TUR, Eq. (\ref{eq:tur_classical}), may be valid if the contribution  
of $\Phi^{qu}_{\alpha}$ is negligible or small compared with that of $\Phi^{cl}_{\alpha}$. 
When the magnitude of  $\Phi^{qu}_{\alpha}$ is prominent, TUR violations are to be observed, that is 
$\sigma(\Phi_{\alpha}^{cl}-\Phi_{\alpha}^{qu})<2$.  
In the next section we explore this scenario in resonant tunnelling junctions.


\section{Resonant Tunnelling transport: conditions for violating the TUR}\label{sec:3}

In this section we derive bounds for the relative uncertainties for 
charge transport, $\sigma\Phi_C^{cl}$ and $\sigma\Phi_C^{qu}$
in the resonant tunnelling regime. 
For clarity, below we separately treat charge transport problems ($\beta_L=\beta_R$ and
$\mu_L\neq \mu_R$), and thermoelectric junctions ($\beta_L\neq\beta_R$ and $\mu_L\neq \mu_R$).

Generally, the current and the noise, Eqs. (\ref{eq:jj}) and (\ref{eq:noise}), have to be evaluated numerically
for a particular form of the transmission function $\mathcal{T}(\epsilon)$.
However, in the resonant tunnelling regime bounds
can be derived without specifying the details of the transmission function.
In this limit, the system-metal coupling is assumed to be weak relative to the temperature,
thus systems' resonances are narrow relative to the width of the Fermi functions.
Considering for simplicity a single, sharp resonance at energy $\epsilon_d$,
the currents and noises become \cite{Ptaszynski.18.PRB}
\begin{eqnarray}
\label{eq:j_resonant}
\langle j_C\rangle_{res} &=& \mathcal{T}_1^0(\tilde{f}_L-\tilde{f}_R), 
\nonumber\\
\langle j_E\rangle_{res}&=& \mathcal{T}_1^1(\tilde{f}_L-\tilde{f}_R),
\nonumber\\
\langle\langle j_C^2\rangle\rangle_{res} &=& \mathcal{T}_1^0[\tilde{f}_L(1-\tilde{f}_R)+\tilde{f}_R(1-\tilde{f}_L)]
-\mathcal{T}_2^0(\tilde{f}_L-\tilde{f}_R)^2,
\nonumber\\
\langle\langle j_E^2\rangle\rangle_{res} &=& 
\mathcal{T}_1^2[\tilde{f}_L(1-\tilde{f}_R)+\tilde{f}_R(1-\tilde{f}_L)]
%
-\mathcal{T}_2^2(\tilde{f}_L-\tilde{f}_R)^2.
\nonumber\\
\end{eqnarray}
The subscript ``res'' highlights that expressions are derived under the resonant tunnelling approximation.
Here, we introduce the coefficients
$\mathcal{T}_n^m\equiv\int_{-\infty}^{\infty}\frac{d\epsilon}{2\pi}\epsilon^m
[\mathcal{T}(\epsilon)]^n$.
The Fermi functions are evaluated at the energy of the resonance, denoted by
 $\tilde{f}_v\equiv f_v(\epsilon_d)$.
In deriving  expressions for the  noise
we replace $f_v(\epsilon)[1-f_v(\epsilon)]=-\beta_v^{-1}\frac{\partial f_v}{\partial \epsilon}$ by $-\beta_v^{-1}\left.\frac{\partial f_v}{\partial \epsilon}\right|_{\epsilon=\epsilon_d}$; 
the first-order derivative of the Fermi distribution is assumed broad relative to the transmission resonance.
While in this section we consider a single resonance of energy $\epsilon_d$,
results can be readily generalized to include multiple states, 
provided that these resonances are sharp and are all located within the thermal window.

\subsection{Charge transport junctions}

We first focus on junctions where $\beta_L=\beta_R=\beta$ and $\mu_L-\mu_R=V$ with $V>0$ 
the applied voltage, that is, we consider a single-affinity charge transfer process.
For simplicity, we further let $\epsilon_d=0$ and set the (equilibrium) Fermi energy at zero. 
Eq. (\ref{eq:j_resonant}) simplifies to
\begin{eqnarray}
\label{eq:j_noise_pct}
\langle j_C\rangle_{res} &=& \mathcal{T}_1^0(\tilde{f}_L-\tilde{f}_R),\nonumber\\
\langle\langle j_C^2\rangle\rangle_{res} &=& \langle\langle j_C^2\rangle\rangle_{res}^{cl}-\langle\langle j_C^2\rangle\rangle_{res}^{qu},
\end{eqnarray}
where $\tilde{f}_v=[e^{-\beta\mu_v}+1]^{-1}$, 
$\langle\langle j_C^2\rangle\rangle_{res}^{cl}=\mathcal{T}_1^0[\tilde{f}_L(1-\tilde{f}_R)+\tilde{f}_R(1-\tilde{f}_L)]$ and $\langle\langle j_C^2\rangle\rangle_{res}^{qu}=\mathcal{T}_2^0(\tilde{f}_L-\tilde{f}_R)^2$. 
%
In arriving at the analytic expressions, Eq. (\ref{eq:j_resonant}), 
one implicitly requires that the transmission function is centered around a single resonance $\epsilon_d$.
It is then reasonable to suggest that $\left.\mathcal{T}(\epsilon)\right|_{\epsilon_d=0}$ 
is an even function of $\epsilon$, and consequently $\mathcal{T}_1^1$ and  $\langle j_E\rangle_{res}$ vanish
 (this should be the case when $\beta_L=\beta_R$). 

In this resonant tunnelling regime, the classical component of the relative uncertainty reduces to
\begin{equation}
\label{eq:cl_tur_pc}
\sigma\Phi_C^{cl}~=~\beta V\coth\left(\frac{\beta V}{2}\right)~\ge~2.
\end{equation}
This result is obtained by noting that the entropy production rate,
$\sigma=\beta V\langle j_C\rangle$ is just the joule's heating. The classical noise is bounded from below by $2$, as expected. Similarly, we find for the quantum component 
$\Phi_C^{qu}$ that ($\Theta\equiv\mathcal{T}_2^0/\mathcal{T}_1^0$ hereafter) 
\begin{eqnarray}\label{eq:qu_tur_pc}
\sigma\Phi_C^{qu} &=& \frac{\beta V}{2}\Theta\left[\tanh\left(\frac{\beta \mu_L}{2}\right)-\tanh\left(\frac{\beta \mu_R}{2}\right)\right]\nonumber\\
&<& \frac{\beta V}{4}\Theta(\beta \mu_L-\beta\mu_R)=\frac{(\beta V)^2}{4}\Theta,
\end{eqnarray}
where we have used the inequality $\tanh x<x$ for $x>0$. 
Combining Eqs. (\ref{eq:cl_tur_pc}) and (\ref{eq:qu_tur_pc}), we get
\begin{equation}\label{eq:bound_pc_resonant}
\sigma\Phi_C-2~>~-\frac{(\beta V)^2}{4}\Theta.
\end{equation}
Since $\beta V$ and $\Theta$ are positive, the above inequality 
indicates that a violation of TUR may occur in the resonant tunnelling regime. 
However, we note that the bound in Eq. (\ref{eq:bound_pc_resonant}) is not a tight one for 
the functional $\sigma\Phi_C-2$,  therefore uninformative.

To be more precise and find whether a violation of TUR can in fact occur, 
we go back to Eqs. (\ref{eq:cl_tur_pc}) and  (\ref{eq:qu_tur_pc}) and combine them.
The explicit functional form of $\sigma\Phi_C-2$ in the resonant tunnelling regime is
\begin{eqnarray}\label{eq:tur_func}
\sigma\Phi_C-2 &=& \beta V\left\{\coth\left(\frac{\beta V}{2}\right)-\frac{\Theta}{2}\left[\tanh\left(\frac{\beta \mu_L}{2}\right)\right.\right.\nonumber\\
&&\left.\left.-\tanh\left(\frac{\beta \mu_R}{2}\right)\right]\right\}-2.
\end{eqnarray}
The voltage may be distributed un-evenly on the two leads, and therefore in the following 
we denote $\mu_L=\frac{\kappa}{1+\kappa}V$ 
and $\mu_R=-\frac{1}{1+\kappa}V$;  $\kappa>0$. 
The case  $\kappa=1$ corresponds to a symmetric bias drop.

Taylor expanding the hyperbolic functions in Eq. (\ref{eq:tur_func}) in powers of $V$, 
the inequality $\sigma\Phi_C-2<0$, measuring the possibility of TUR violations, translates to
\begin{equation}\label{eq:expansion}
(\beta V)^4\left(\frac{\Theta}{24}K-\frac{1}{180}\right)-(\beta V)^2\left(\frac{\Theta}{2}-\frac{1}{3}\right)~<~0,
\end{equation}
where we have kept terms up to the order of $\mathcal{O}(V^4)$. 
We denoted by $K=\frac{\kappa^2-\kappa+1}{(\kappa+1)^2}$;
note that $\frac{1}{4}\leq K\leq1$.
We now identify the range for voltage in which the above inequality holds,
\begin{equation}\label{eq:bias_range}
0~<~V~<~V^{\ast}=\frac{4}{\beta}\sqrt{\frac{15(3\Theta-2)}{60K\Theta-8}},
\end{equation}
provided that we also satisfy $\Theta>\frac{2}{3}$. Remarkably, this  condition on the ratio $\Theta$ 
was obtained in Ref. \cite{Bijay.18.PRB} as a result of the fluctuation symmetry.  

It can be verified that $V^{\ast}$ reaches its maximum value 
$V_{max}^{\ast}$
when $\kappa=1$ ($K=\frac{1}{4}$) and its minimum value $V_{min}^{\ast}$
when $\kappa=0$ or $\kappa\to\infty$ ($K=1$). 
This result can be understood by noting that the setup corresponding to either 
$\kappa=0$ or $\kappa\to\infty$ favors single particle transfer processes 
and thus the impact of the quantum component $\langle\langle j_C^2\rangle\rangle_{qu}$ is minimal.
The contribution of $\langle\langle j_C^2\rangle\rangle_{qu}$ is most significant in a
symmetric setup corresponding to $\kappa=1$. 
This sensitivity of the current noise to the bias splitting
thus allows us to enhance or suppress it by simply adjusting the partitioning of the
chemical potentials of metal leads.

In conclusion, in accord with Ref. \cite{Bijay.18.PRB}, charge transport junctions can violate the TUR
in a certain range of voltage, which should not exceed few $k_BT$.
The new result, Eq. (\ref{eq:bias_range}) allows us to identify the range of TUR violations
given the junctions' parameters, $\mathcal T_{1,2}^0$, and the thermodynamic-external variables,
 $\kappa$ and $\beta$. 

\subsection{Thermoelectric engines}
We now turn to junctions that operate as thermoelectric engines:
we let $\beta_L>\beta_R$, $\mu_L>\mu_R$ such that charge and energy currents can be driven against 
the voltage due to the temperature gradient. 
The junction operates as a thermoelectric engine (power is produced) 
when both charge and energy current flow from the 
hot (right) lead to the cold one, against the applied voltage.
The average power done by the engine then reads 
$\dot{W}\equiv\langle j_p\rangle=-(\mu_L-\mu_R)\langle j_C\rangle$ 
and the average heat current from the hot (right) lead to the engine is given by 
$\dot{Q}\equiv\langle j_q\rangle=-(\langle j_E\rangle-\mu_R\langle j_C\rangle)$;
recall our convention that currents are positive when flowing from the left lead to the right one. 
The efficiency of such a thermoelectric engine is thus defined as 
$\eta=\langle j_p\rangle/\langle j_q\rangle$. Using Eq. (\ref{eq:sigma_ge}), we
write the entropy production as
\begin{equation}
\sigma~=~\langle j_p\rangle\beta_L\frac{\eta_c-\eta}{\eta},
\label{eq:sigma}
\end{equation}
with $\eta_c=1-\beta_R/\beta_L$ the Carnot efficiency. 

We set $\epsilon_d\neq0$, such that 
$\mathcal{T}_1^1=\epsilon_d\mathcal{T}_1^0\neq0$ in Eq. (\ref{eq:j_resonant}). 
In particular in the resonant tunnelling regime we 
require that $\tilde{f}_L<\tilde{f}_R$ as currents should be negative, or equivalently,
\begin{equation}\label{eq:cc}
-\mathcal{F}_E\epsilon_d>\mathcal{F}_C.
\end{equation}
The efficiency cannot exceed unity, $\eta=(\mu_L-\mu_R)/(\epsilon_d-\mu_R)\leq1$,
therefore $\epsilon_d>\mu_L$.
For later convenience, we introduce the notation
\bea
Y\equiv\mathcal{F}_E\epsilon_d+\mathcal{F}_C.
\label{eq:Y}
\eea
The fact that $Y<0$ when the junction operates as a 
thermoelectric engine, Eq. (\ref{eq:cc}), 
will become critical in our analysis of the possible violation of the TUR.

Here we only perform our analytic study based on the relative uncertainty for charge transport,
$\sigma\Phi_C$;  we have checked that $\sigma\Phi_E$ leads to the same conclusions regarding the 
validity of TUR, as expected in the resonant regime. 
For its classical part, using Eqs. (\ref{eq:j_resonant}) we have
\begin{equation}\label{eq:cl_tur_te}
\sigma\Phi_C^{cl}~=~Y\coth\left(\frac{Y}{2}\right)~\ge~2,
\end{equation}
which is still bounded from below by 2, as expected. Similarly, we find
\begin{eqnarray}\label{eq:qu_tur_te}
\sigma\Phi_C^{qu} &=& Y\frac{\Theta}{2}\left[\tanh\frac{\beta_R(\epsilon_d-\mu_R)}{2}-\tanh\frac{\beta_L(\epsilon_d-\mu_L)}{2}\right]\nonumber\\
&<& Y^2\frac{\Theta}{4},
\end{eqnarray}
as a result of $\tanh x<x$ for $x>0$. 
We now combine Eqs. (\ref{eq:cl_tur_te}) and (\ref{eq:qu_tur_te}) to construct the total relative uncertainty,
$\sigma\Phi_C= \sigma\Phi_C^{cl}-\sigma\Phi_C^{qu}$, and get
\begin{equation}\label{eq:bound_te_resonant}
\sigma\Phi_C-2~>~-Y^2\frac{\Theta}{4}.
\end{equation}
Nevertheless, the above bound is not tight, and it may overestimate the magnitude of 
current noises. Therefore, it cannot conclusively indicate whether a violation of TUR can occur. 

For a definite answer (in the resonant tunnelling regime), we consider the functional
%
\begin{eqnarray}\label{eq:tur_func_2}
\sigma\Phi_C-2 &=& Y\left[\coth\frac{Y}{2}-\frac{\Theta}{2}\left(\tanh\frac{\beta_R(\epsilon_d-\mu_R)}{2}\right.\right.\nonumber\\
&&\left.\left.-\tanh\frac{\beta_L(\epsilon_d-\mu_L)}{2}\right)\right]-2,
\end{eqnarray}
where we combined the quantum and classical uncertainties.
By expanding the hyperbolic functions in powers of chemical potentials
 and keeping terms up to the fourth order, 
we find that violation of TUR, that is, $\sigma\Phi_C-2<0$, is quantified by
\begin{equation}\label{eq:tur_ineq}
\left(15\Theta-2\right)Y^2-45\beta_R\epsilon_d\Theta Y~<~(180-45\beta_R^2\epsilon_d^2)\Theta-120.
\end{equation}
For simplicity, here we set $\mu_L=V$ and $\mu_R=0$,  and already
omitted the common factor $Y^2$, as it is nonzero. 
Eq. (\ref{eq:tur_ineq}) reduces 
to Eq. (\ref{eq:expansion}) with $K=1$ once we set $\epsilon_d=0$, $\beta=\beta_v$, 
with  the thermodynamic force $Y=\beta V$.
The inequality (\ref{eq:tur_ineq}) is solved in the Appendix, identifying the possible range of 
the variable $Y$ for achieving TUR violation.

The ratio $\Theta\equiv\mathcal{T}_2^0/\mathcal{T}_1^0$  
is an intrinsic property of the system 
irrespective of temperatures and chemical potentials. As well,  
$\Theta$ does not depend on the resonance energy $\epsilon_d$ since the integral
$\int_{-\infty}^{\infty} 
\mathcal [T(\epsilon)]^n d\epsilon$ can be shifted around $\epsilon-\epsilon_d$.
This implies that to observe TUR violation in a thermoelectric junction, 
we should still enforce $\Theta>2/3$ as in the charge transport system 
(a detailed proof is given in the Appendix).

We are interested in satisfying Eq. (\ref{eq:tur_ineq}), that is in breaking the TUR,
{\it while  producing output power} ($Y<0$).
In the Appendix we show that Eq. (\ref{eq:tur_ineq}) is satisfied,
with a negative value $Y$, when
\begin{equation}\label{eq:br_ed}
0~<~\beta_R^2\epsilon_d^2~<~4-\frac{8}{3\Theta}.
\end{equation}
It is significant to note that $\Theta$ cannot be arbitrarily large.
In fact for the serial DQD system discussed in the next section, $0<\Theta<0.782$,
limiting the range of parameters that permit TUR violations along with 
power generation. Crucially, simulations in Sec. \ref{sec:4} illustrate that while thermoelectric
generators can violate the TUR at finite power when $\eta<\eta_c$,
the TUR is recovered once we approach the maximum (Carnot) efficiency.

It was recently claimed in Ref. \cite{Ptaszynski.18.PRB} that one could have a coherence-enhanced 
constancy (reduced noise) for thermoelectric engines in the resonant tunnelling regime, 
as a consequence of the violation 
of TUR. However, we point out that the system considered in that paper in fact did not
operate as a thermoelectric engine since it was studied for $T_L>T_R$ and $\mu_L>\mu_R$.

Although the above analysis is based on expressions from the resonant tunnelling regime, 
one can argue that violating the TUR in the strong system-bath coupling regime is highly unlikely:
First, in the extreme limit of $\mathcal{F}_E=0$, thermoelectric engines reduce to pure charge transfer junction.
However, we know that we cannot violate the TUR in the strong coupling regime 
for a serial quantum dots \cite{Bijay.18.PRB}.  From the other end, in the case of a single quantum dot 
the TUR can be violated for a pure charge transport at strong coupling, but
in this regime the system does not act as a thermoelectric generator.
Exact simulations below provide evidences to support our argument that noninteracting-electron quantum 
dot thermoelectric generators satisfy the TUR in the strong coupling regime.


\section{Case study: Serial quantum dot}\label{sec:4}

To verify and assess our theoretical results, we consider a serial DQD junction. 
The model consists of two interacting quantum dots of energies $\epsilon_L$ and $\epsilon_R$, with 
the tunnelling element $\Omega$, coupled in series to two leads. 
The transmission function is given by \cite{Hartle.13.PRB,Simine.15.JPCC,Bijay.18.PRB,Ptaszynski.18.PRB}
\begin{equation}
\mathcal{T}(\epsilon)~=~\frac{\Gamma_L\Gamma_R\Omega^2}{|(\epsilon-\epsilon_L+i\Gamma_L/2)(\epsilon-\epsilon_R+i\Gamma_R/2)-\Omega^2|^2}.
\end{equation}
Assuming a symmetric coupling $\Gamma_L=\Gamma_R=\Gamma$ and degenerate orbital energies 
$\epsilon_L=\epsilon_R=\epsilon_d$, the transmission function is simplified to
\begin{equation}\label{eq:trans}
\mathcal{T}(\epsilon)~=~\frac{\Gamma^2\Omega^2}{\left[(\epsilon-\epsilon_d+\Omega)^2+\frac{\Gamma^2}{4}\right]\left[(\epsilon-\epsilon_d-\Omega)^2+\frac{\Gamma^2}{4}\right]}.
\end{equation}
Inserting Eq. (\ref{eq:trans}) into Eqs. (\ref{eq:jj}) and (\ref{eq:noise}), then performing numerical integration, 
we easily obtain exact numerical results for the DQD system. 
We further test the applicability of the resonant tunneling expressions of Sec. \ref{sec:3}, 
and therefore calculate 
the coefficients in Eq. (\ref{eq:j_resonant}),
\begin{eqnarray}\label{eq:t12}
\mathcal{T}_1^0 &=& \frac{2\Gamma\Omega^2}{\Gamma^2+4\Omega^2},\nonumber\\
\mathcal{T}_1^1 &=& \frac{2\epsilon_d\Gamma\Omega^2}{\Gamma^2+4\Omega^2},\nonumber\\
\mathcal{T}_1^2 &=& \frac{\Gamma\Omega^2[\Gamma^2+4(\epsilon_d^2+\Omega^2)]}{2(\Gamma^2+4\Omega^2)},\nonumber\\
\mathcal{T}_2^0 &=& \frac{4\Gamma\Omega^4(5\Gamma^2+4\Omega^2)}{(\Gamma^2+4\Omega^2)^3},\nonumber\\
\mathcal{T}_2^2 &=& \frac{\Gamma\Omega^4[\Gamma^4+4\Gamma^2(5\epsilon_d^2+2\Omega^2)+16\Omega^2(\epsilon_d^2+\Omega^2)]}{(\Gamma^2+4\Omega^2)^3}.
\nonumber\\
\end{eqnarray}
In what follows we perform numerical simulations for the charge and energy currents, 
their noises and the combination $\sigma \Phi_{C,E}-2$, which when negative establishes TUR violations,
in both charge transport and thermoelectric junctions.

\subsection{Charge transport junctions}

We consider the relative uncertainty (\ref{eq:tur_func}) and evaluate it using
$\mathcal{T}_1^0$ and $\mathcal{T}_2^0$ from Eq. (\ref{eq:t12}).
We further define the ratio $R\equiv\Gamma/\Omega$ and organize
\begin{eqnarray}\label{eq:tur_func_s}
\frac{\sigma\Phi_C}{\beta V} &=& \coth\left(\frac{\beta V}{2}\right)-\frac{(5R^2+4)}{(R^2+4)^2}\left[\tanh\left(\frac{\kappa\beta V}{2(1+\kappa)}\right)\right.\nonumber\\
&&\left.+\tanh\left(\frac{\beta V}{2(1+\kappa)}\right)\right]. 
\end{eqnarray}
The TUR can be violated in the voltage range described by Eq. (\ref{eq:bias_range}), with the maximal voltage,
\begin{equation}\label{eq:v_ast_dqd}
V^{\ast}~=~\frac{4\sqrt{15}}{\beta}\sqrt{\frac{12(5R^2+4)-4(R^2+4)^2}{240K(5R^2+4)-16(R^2+4)^2}}.
\end{equation}
These two expressions, Eqs. (\ref{eq:tur_func_s}) and (\ref{eq:v_ast_dqd}) immediately 
lead to some interesting  observations.
For a specific configuration in which $\kappa$ is fixed, if we further set the ratio $R$, 
$\sigma\Phi_C-2$ is simply a function of a scaled voltage $\beta V$. Further,
the voltage range $0<V<V^{\ast}$ in which the violation of TUR occurs depends only on the inverse temperature 
$\beta$ of metal leads (for a fixed $R$). 
When the inverse temperature $\beta$ is fixed, $V^{\ast}$ is solely determined by the ratio $R$. 
We therefore expect to observe the violation of TUR within the same voltage range for setups with different values 
of $\Gamma$ and $\Omega$, as long as they build the same ratio $R$ (again, assuming
the resonant tunnelling regime). In particular, $\sigma\Phi_C-2$ should collapse into a single curve when varying $V$.

It should be pointed out that
the condition $\Theta>\frac{2}{3}$, which is necessary for TUR violation, 
constrains the coupling strength, 
$\Omega\sqrt{\frac{7-\sqrt{33}}{2}}~<~\Gamma~<~\Omega\sqrt{\frac{7+\sqrt{33}}{2}}$ for fixed $\Omega$,
 or $\Gamma\sqrt{\frac{7-\sqrt{33}}{8}}~<~\Omega~<~\Gamma\sqrt{\frac{7+\sqrt{33}}{8}}$ for fixed $\Gamma$. 
This condition can be also organized for the ratio $R=\Gamma/\Omega$, as $0.79<R<2.52$ 
\cite{Ptaszynski.18.PRB,Bijay.18.PRB}.

\begin{figure}[tbh!]
  \centering
  \includegraphics[width=0.85\columnwidth]{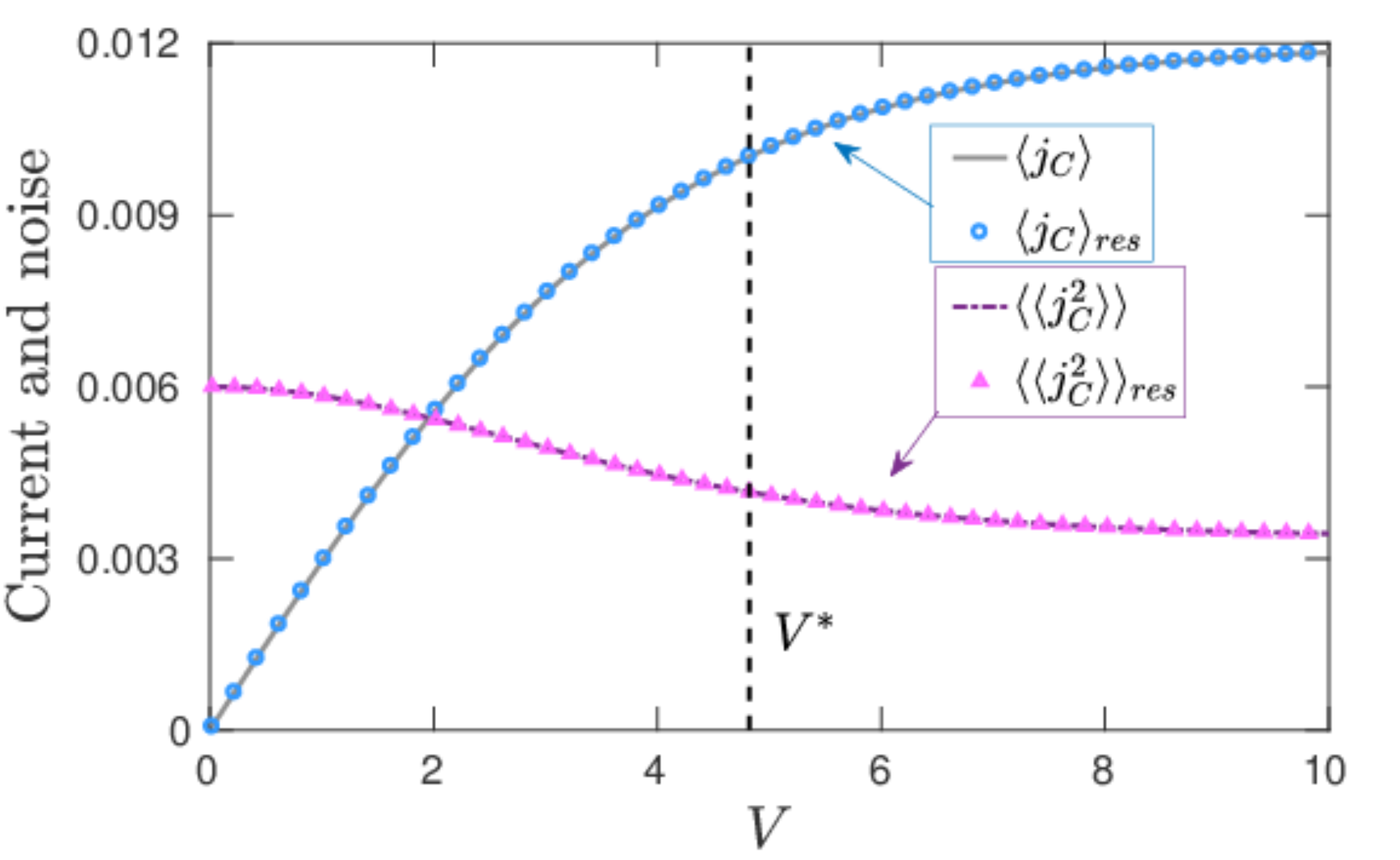} 
\caption{Charge current and its noise as a function of voltage $V$. 
Exact results for $\langle j_C\rangle$ (solid) and $\langle\langle j_C^2\rangle\rangle$ (dashed-dotted)
are compared to the resonant tunneling expressions (\ref{eq:j_resonant}),
$\langle j_C\rangle_{res}$ ($\circ$) and $\langle\langle j_C^2\rangle\rangle_{res}$ ($\triangle$).
The vertical black dashed line denotes the location of $V^{\ast}$. 
The values of parameters are $\beta=1$, $\Gamma=\Omega=0.03$, $\mu_L=V/2$, $\mu_R=-V/2$ and $\epsilon_d=0$.} 
\label{fig:cur_noise}
\end{figure}

The above theoretical perspectives were gained based on analytic expressions in the resonant tunnelling regime.
In Fig. \ref{fig:cur_noise}, we compare analytic results for charge current and noise 
[Eqs. (\ref{eq:j_resonant}) with (\ref{eq:t12})]
to exact simulations. 
By choosing $\Omega$, $\Gamma\ll\beta^{-1}$,  we find that the agreement between analytic predictions (symbols) and exact numerical results (lines) is excellent,
thereby confirming the validity of Eq. (\ref{eq:j_resonant}) and consequently the above theoretical analysis 
in the resonant tunnelling regime.

\begin{figure}[tbh!]
 \centering
 \includegraphics[width=1.05\columnwidth]{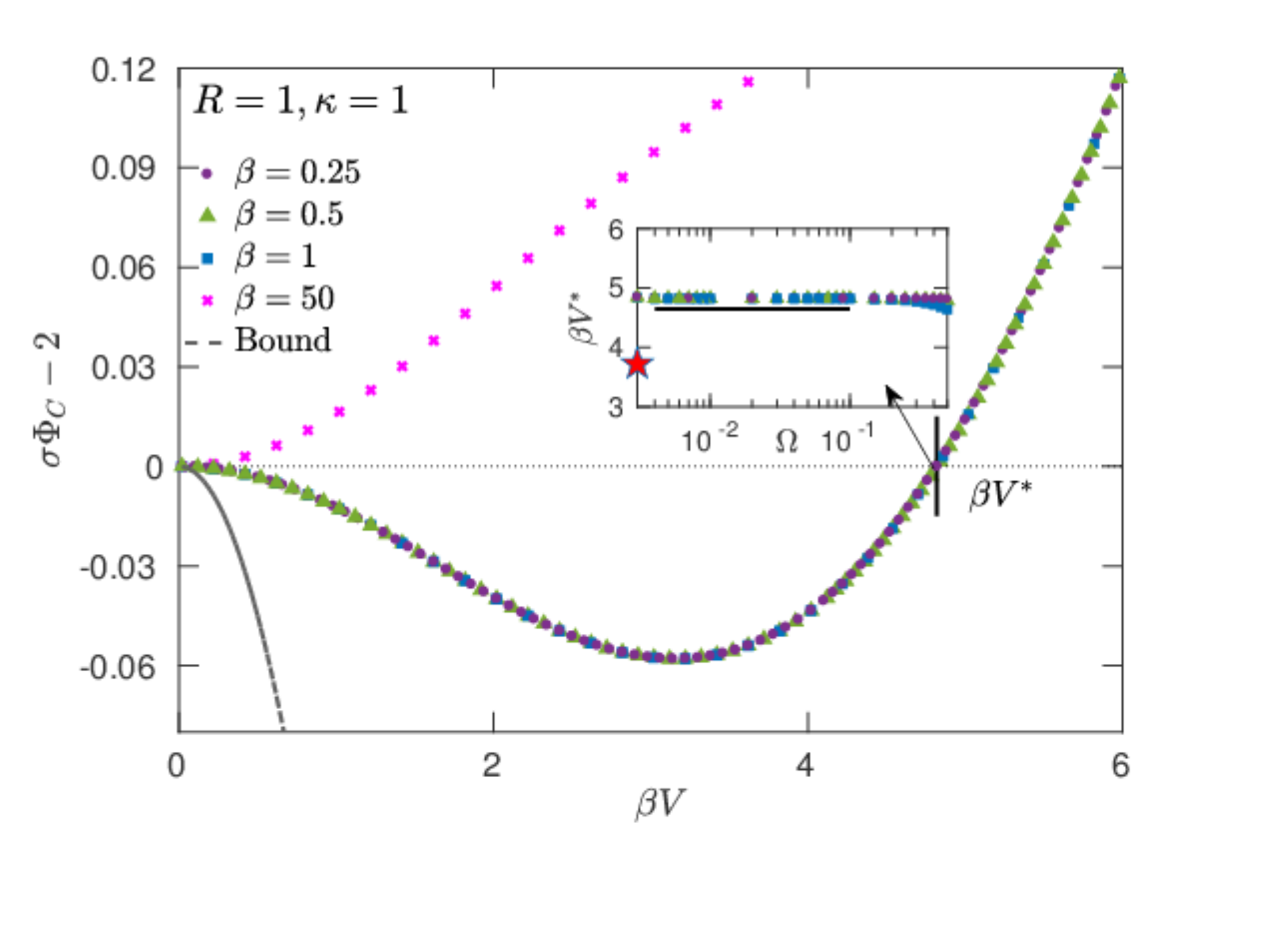}  
\caption{Exact simulations for $\sigma\Phi_C-2$ as a function of scaled voltage $\beta V$ 
for $\beta=0.25$ (circle), $\beta=0.5$ (triangle), $\beta=1$ (square), and $\beta=50$ (x).
 We fix $R=1$ with $\Gamma=\Omega=0.03$.
The dashed line corresponds to the non-tight bound calculated from Eq. (\ref{eq:bound_pc_resonant}). 
(inset) We vary  $\Omega$ ($R=1$ is fixed) and
 extract $\beta V^{\ast}$ at the (nonequilibrium) point
 $\sigma\Phi_C=2$.
The red star on the y-axis marks the analytic result, $\beta V^{\ast}\approx 3.7$, 
from Eq.~(\ref{eq:v_ast_dqd}). 
%
The horizontal solid line in the inset serves as  a guide to the eye. 
The values of the remaining parameters are $\kappa=1$, 
corresponding to symmetric splitting $\mu_L=V/2$ and $\mu_R=-V/2$, $\epsilon_d=0$.} 
\label{fig:tur_1}
\end{figure}
%

In Fig. \ref{fig:tur_1} we fix $R=1$ with $\Omega=\Gamma=0.03$. 
As can be seen, in the resonant tunneling regime when $\beta^{-1}\gg\Omega, \Gamma$,
$\sigma\Phi_C-2$ is solely determined by the scaled voltage $\beta V$.
In the inset, we compare $\beta V^{\ast}$ for different temperatures by varying the values of $\Gamma$ and $\Omega$ 
while maintaining the ratio, $R=1$. 
It is evident that $\beta V^{\ast}$ reaches a constant value 
in the resonant tunnelling regime, and begins to 
show a $\Gamma$-dependence in the intermediate coupling regime. 
If we further increase the inverse temperature $\beta$, 
the system eventually enters into a strong coupling regime. 
Specifically, we demonstrate that  for $\beta=50$ the TUR is always valid with  $\sigma\Phi_C\geq2$.
In fact, $V^{\ast}\to0$ when $\beta\to\infty$ according to Eq. (\ref{eq:v_ast_dqd}). 

These observations hold for other values of $R=\Gamma/\Omega$ in the appropriate range
(remember that the condition for TUR violation,
$\Theta>\frac{2}{3}$ enforces
$0.79<R<2.52$ for the serial double dot junction).
For example, we confirm with exact simulations (not shown) that when varying $\Omega$ 
at $R=1.6$, 
the curves for $\sigma\Phi_C-2$ coincide with each other (as long as $\Omega$, $\Gamma$ $<\beta$),
irrespective of the actual values of $\Gamma$ and $\Omega$. 

Although exact simulations confirm our theoretical predictions, 
Eqs. (\ref{eq:tur_func_s}) and (\ref{eq:v_ast_dqd}), we notice that the analytic value for $V^{\ast}$ 
obtained from Eq. (\ref{eq:v_ast_dqd}) deviates from the exact counterpart (red star in the y-axis 
of the inset), implying the importance of higher-order terms that 
we neglected in obtaining Eq. (\ref{eq:expansion}). 
Nevertheless, Eqs. (\ref{eq:tur_func_s}) and (\ref{eq:v_ast_dqd}) capture the essential basic physical features. 
Moreover, we note that the bound of Eq. (\ref{eq:bound_pc_resonant}) 
is rather loose and it drops very quickly with voltage.

\begin{figure}[tbh!]
  \centering
  \includegraphics[width=1.05\columnwidth]{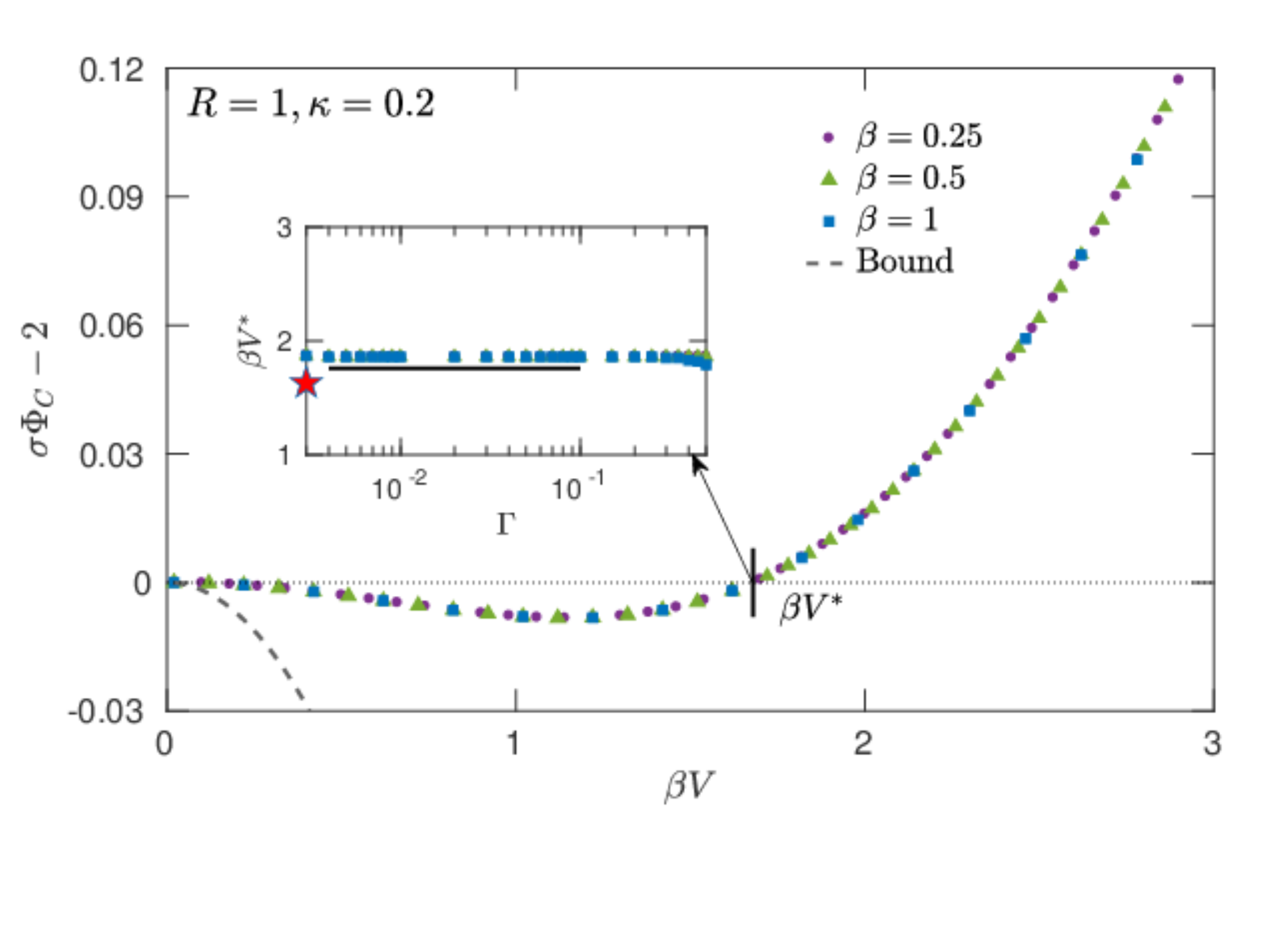}   
\caption{Exact simulations for $\sigma\Phi_C-2$ as a function of scaled voltage $\beta V$ 
for $\beta=0.25$ (circle), $\beta=0.5$ (triangle), and $\beta=1$ (square)
for an asymmetric potential drop, $\kappa=0.2$.
We fix $R=1$ with $\Gamma=\Omega=0.05$. 
The dashed line is the bound (\ref{eq:bound_pc_resonant}). 
(inset) We extract the value of $\beta V^{\ast}$ for the same temperatures while
 varying $\Gamma$ ($R=1$ is fixed). 
The red star on the y-axis corresponds to the analytic result
from Eq.~(\ref{eq:v_ast_dqd}). 
The horizontal solid line in the inset serves as a guide to the eye. 
The values of the remaining parameter is $\epsilon_d=0$.}
\label{fig:tur_3}
\end{figure}

In Fig. \ref{fig:tur_3} we study the role of bias voltage asymmetry, $\kappa\neq1$.
In particular, we use $\mu_L=V/6$ and $\mu_R=-5V/6$, which corresponds to $\kappa=0.2$.
As expected, the curve $\sigma\Phi_C-2$ 
(for different inverse temperatures) still collapses 
into a single curve by scaling the voltage with the corresponding temperature. 
These results further indicate the utility of analytic expressions,
 Eqs. (\ref{eq:tur_func_s}) and (\ref{eq:v_ast_dqd}) in capturing essential physics. 
Similarly, the bound Eq. (\ref{eq:bound_pc_resonant}) is correct, 
although it is not tight (therefore not very useful).

\begin{figure}[tbh!]
  \centering
  \includegraphics[width=1.05\columnwidth]{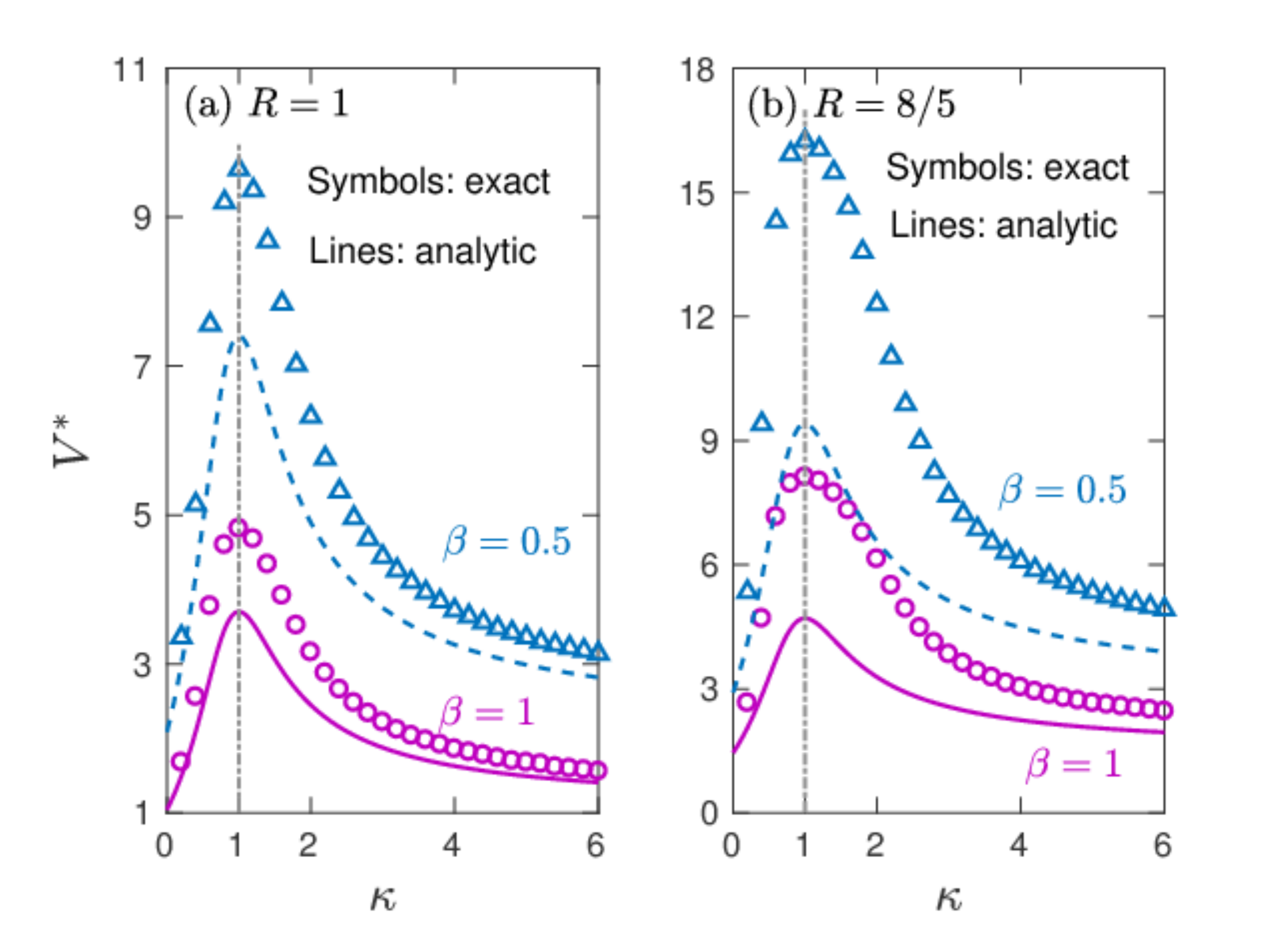}   
\caption{$V^{\ast}$ as a function of $\kappa$ for different dot-lead and inter-dot coupling strength ratios, 
(a) $R=1$ with $\Gamma=\Omega=0.005$, (b) $R=8/5$ with $\Gamma=0.008$ and $\Omega=0.005$. 
In both plots, exact simulations and analytic predictions from Eq. (\ref{eq:bias_range}) 
are denoted by symbols (triangles and circles) and lines (dashed and solid), respectively. 
Results for $\beta=0.5$ and $\beta=1$ are identified in the figure.
The vertical dashed-dotted line corresponds to $\kappa=1$.} 
\label{fig:tur_2}
\end{figure}

To understand TUR violation, it is critical to identify the range of voltage where it takes place.
In Fig. \ref{fig:tur_2} we study the behavior of $V^{\ast}$ by varying the partitioning of the
chemical potential $\kappa$. 
We consider two coupling strength ratios, $R=1$ and $R=8/5$, see panels (a) and (b), respectively,
with appropriate values of $\Gamma$ and $\Omega$.
In both panels, we see that exact results for $V^{\ast}$ reach maximum at $\kappa=1$,
and decrease when $\kappa$ shifts away from 1. 
The analytic expression of Eq. (\ref{eq:bias_range}), although deviates from exact results, 
qualitatively captures the trends for $V^{\ast}$. 
In principle, this trend implies that we can tune charge current fluctuations by simply 
adjusting the fraction of potential drop at the metals.

%

We conclude this section:
(i) The TUR can be violated in charge conducting junctions within a certain range of voltage and $\Gamma/\Omega$.
(ii) At high enough voltage dissipation is significant and the TUR is obeyed.
(iii) TUR violation can be controlled by adjusting $\kappa$, the partitioning of voltage in the leads.

\subsection{Thermoelectric engines}

In this section we investigate the TUR in the serial DQD model, focusing
on the regime where it operates as a thermoelectric generator.
In Sec. \ref{sec:3}, we showed that in the resonant tunnelling regime
whether the TUR holds in thermoelectric engines depends on the value of $\beta_R\epsilon_d$.
Here, we (i)  test this prediction with exact numerical simulations,
(ii)  analyze the system beyond the resonant tunnelling regime, 
(ii) demonstrate that while thermoelectric power generators can violate the TUR
within a certain range of parameters,
the TUR is recovered as we approach the Carnot efficiency, satisfying in this limit
a trade-off relation between power production, efficiency and power fluctuations.

\begin{figure}[tbh!]
  \centering
  \includegraphics[width=1.05\columnwidth]{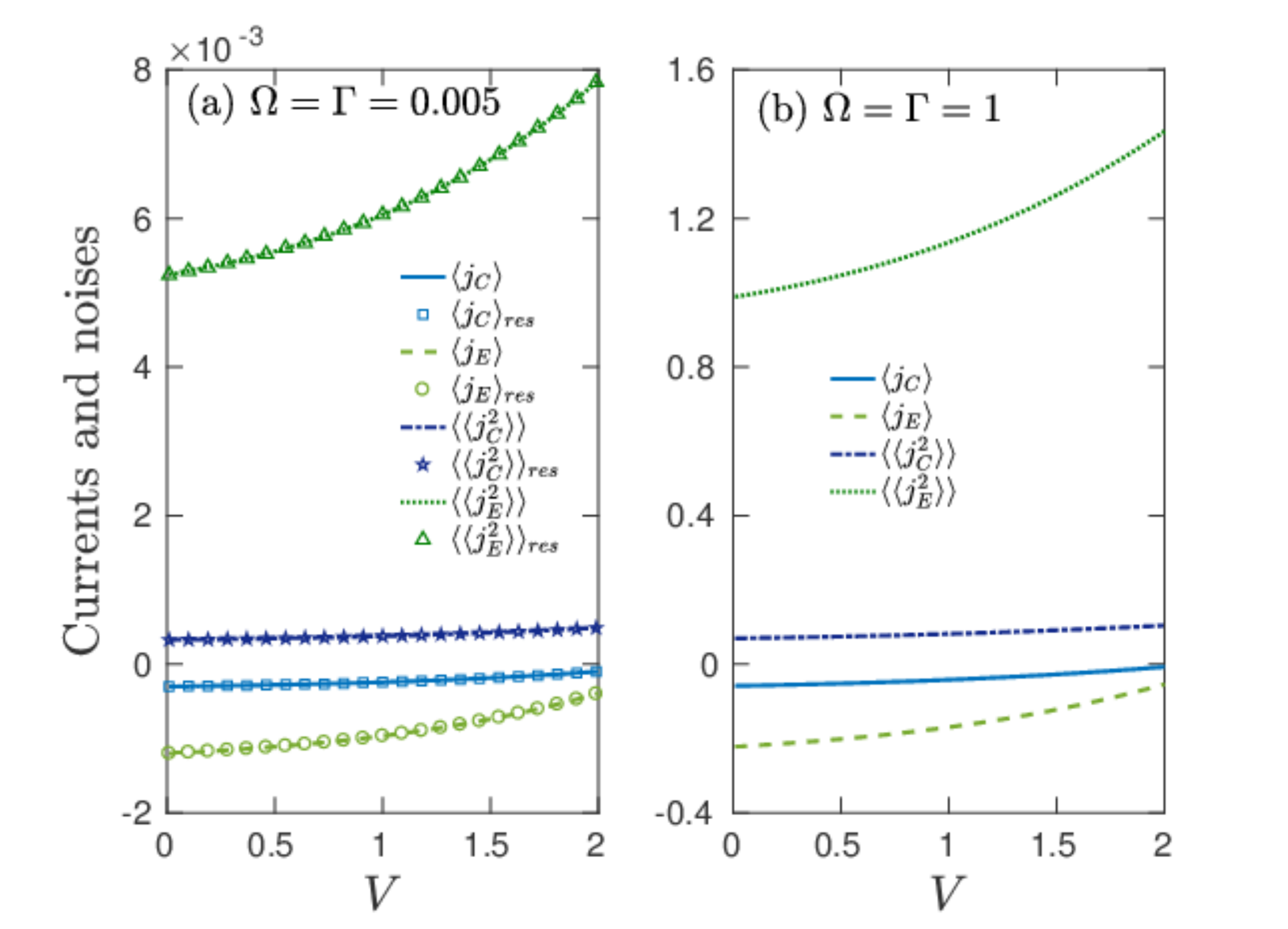}  
\caption{Results for currents $\langle j_C\rangle$, $\langle j_E\rangle$, 
and noises $\langle\langle j_C^2\rangle\rangle$, $\langle\langle j_E^2\rangle\rangle$ 
as a function of voltage $V$ while 
varying $\Omega$ and $\Gamma$. 
(a) Resonant tunneling regime, $\Omega=\Gamma=0.005$. 
(b) Intermediate coupling regime, $\Omega=\Gamma=1$. 
In both panels, exact results are identified by lines.
In (a), analytic results obtained from Eq. (\ref{eq:j_resonant}) are denoted by symbols.
The values of the remaining parameters 
are $\beta_L=1$, $\beta_R=0.4$, $\mu_L=V$, $\mu_R=0$, $\epsilon_d=4$.} 
\label{fig:cur_te}
\end{figure}

First, in Fig.  \ref{fig:cur_te}, we illustrate the behaviors of currents and noises in 
different coupling regimes.
Both charge and energy currents are negative, following our sign convention,
implying that both currents flow from the right to the left lead,
and that the thermoelectric junction indeed operates as a thermoelectric engine 
with the parameters we select.
Since $\beta_L\neq\beta_R$, there are finite currents even when $V=0$. 
For comparison,  in Fig. \ref{fig:cur_te} (a) we also
present analytic results obtained by 
inserting Eq. (\ref{eq:t12}) into Eq. (\ref{eq:j_resonant}). 
As expected, these analytic results agree very well with exact simulations in the 
resonant tunnelling regime.

\begin{figure}[tbh!]
  \centering
  \includegraphics[width=1.05\columnwidth] {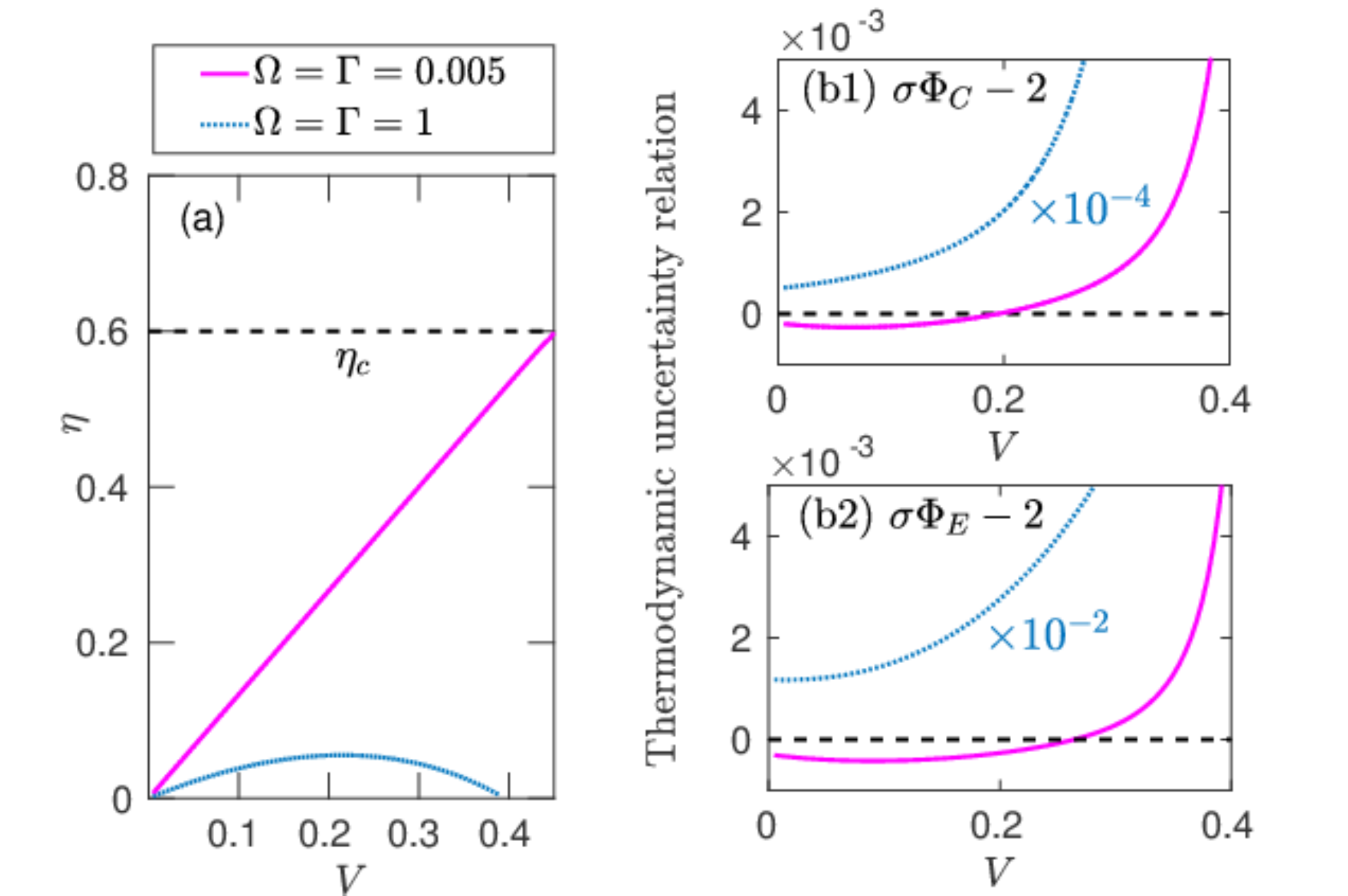} 
\caption{(a) The efficiency $\eta$ of a thermoelectric generator as a function of voltage
based on exact simulations with
$\Omega=\Gamma=0.005$ (solid line) and $\Omega=\Gamma=1$ (dotted line).
The horizontal dashed line marks the Carnot efficiency, $\eta_c=1-\beta_R/\beta_L$.
(b1)-(b2) Corresponding exact simulation for  $\sigma\Phi_C-2$ and $\sigma\Phi_E-2$, respectively.
The values of the remaining parameters are 
$\beta_L=1$, $\beta_R=0.4$, $\mu_L=V$, $\mu_R=0$, $\epsilon_d=0.75$.
Since $\epsilon_d$ complies with Eq. (\ref{eq:br_ed}), the system realizes 
TUR violation at finite output power---yet below the Carnot efficiency bound.
}
\label{fig:tur_te_small}
\end{figure}


\begin{figure}[tbh!]
\centering
  \includegraphics[width=1.05\columnwidth]{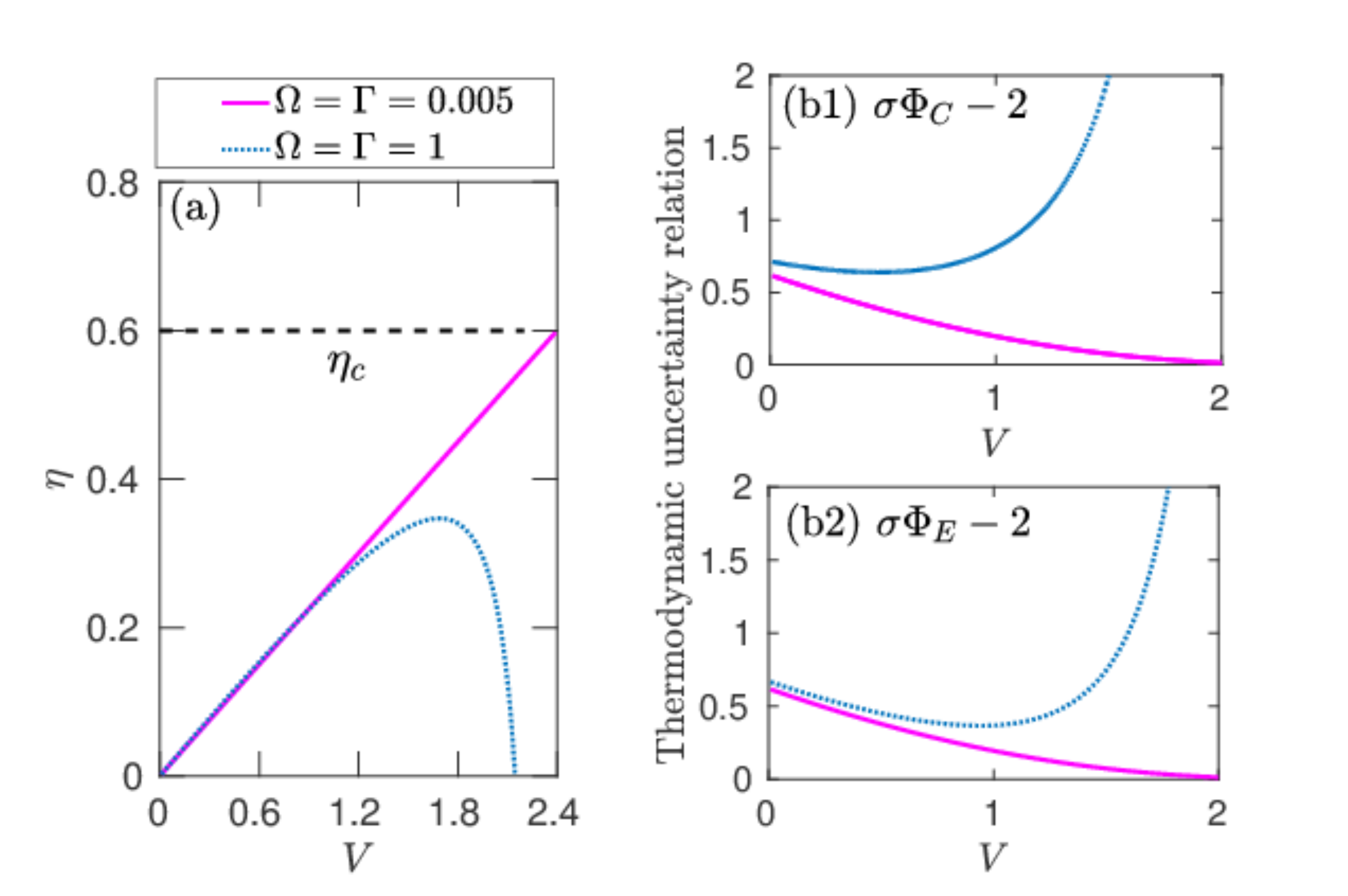}   
\caption{(a) The efficiency $\eta$ of a thermoelectric generator as a function of voltage
based on exact simulations with
$\Omega=\Gamma=0.005$ (solid line) and $\Omega=\Gamma=1$ (dotted line).
The horizontal dashed line marks the Carnot efficiency, $\eta_c=1-\beta_R/\beta_L$.
(b1)-(b2) Corresponding exact simulation for  $\sigma\Phi_C-2$ and $\sigma\Phi_E-2$, respectively.
The values of the remaining parameters are 
$\beta_L=1$, $\beta_R=0.4$, $\mu_L=V$, $\mu_R=0$, $\epsilon_d=4$.
}
\label{fig:tur_te}
\end{figure}
\begin{figure}[tbh!]
\hspace{-15mm}
\includegraphics[width=1.15\columnwidth] {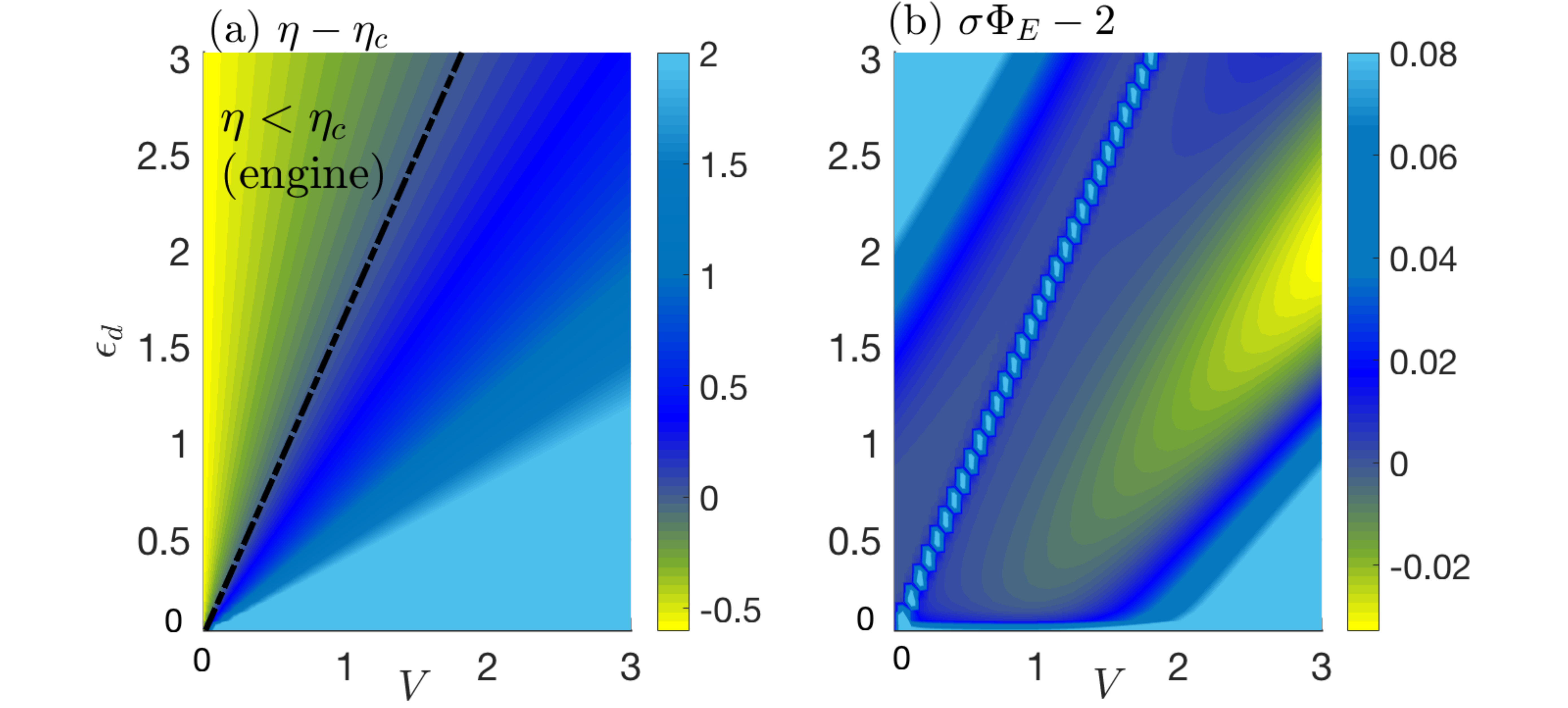}
\caption{Contour map of (a) $\eta-\eta_c$ 
and (b) $\sigma\Phi_E-2$ as a function of voltage and dot energy.
The  dashed line in (a) marks the Carnot efficiency, $\eta=\eta_c$.
The system does not operate as an engine to the right of this line.
The values of the remaining parameters are $\beta_L=1$, $\beta_R=0.4$, $\mu_L=V$, $\mu_R=0$, and $\Gamma=\Omega=0.005$.}
\label{fig:con}
\end{figure}

\begin{figure}[tbh!]
  \centering
  \includegraphics[width=1.05\columnwidth]{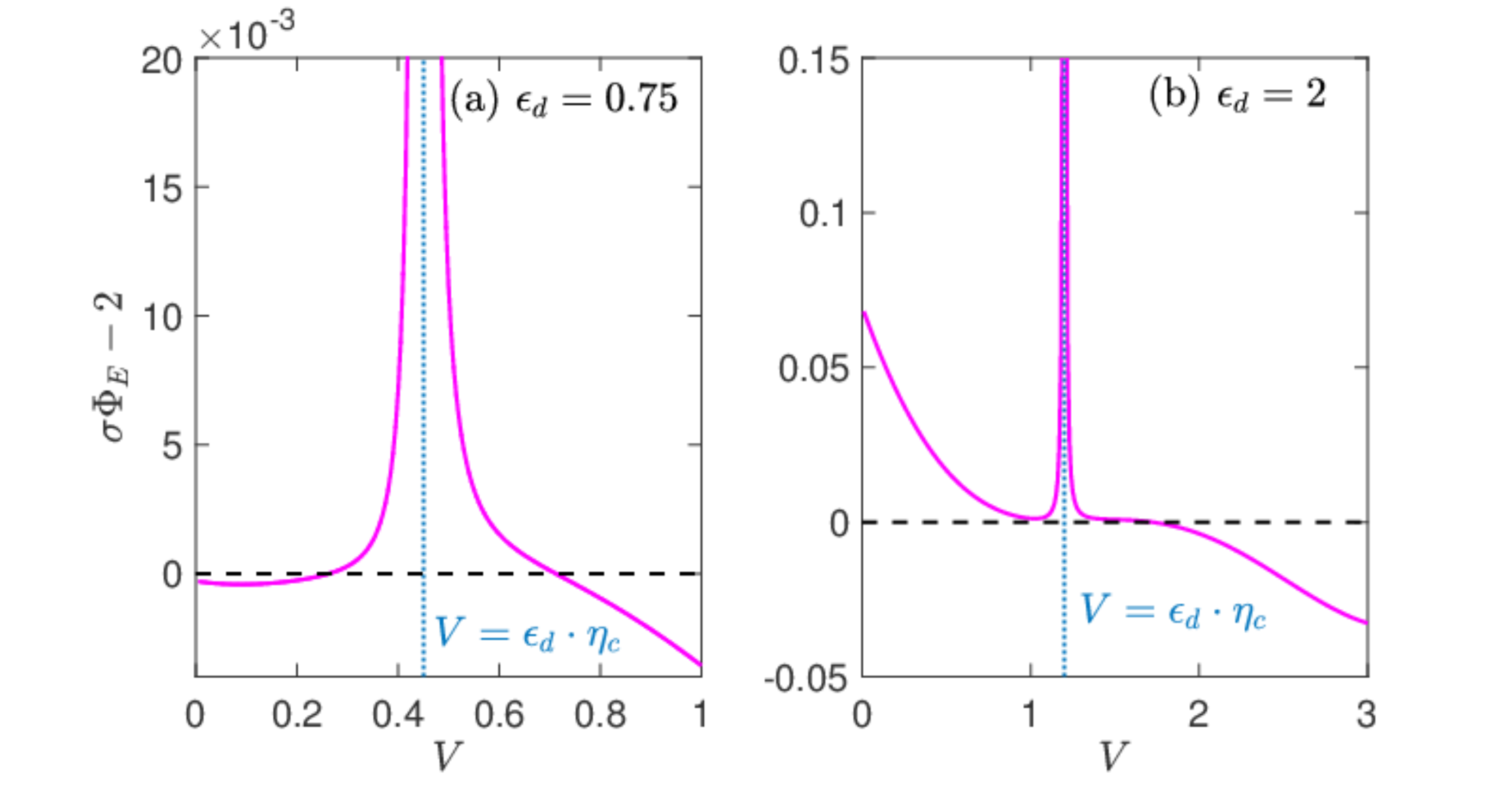}
\caption{Divergence of $\sigma\Phi_E-2$ when approaching the maximal (Carnot) efficiency
based on exact simulations with (a) $\epsilon_d=0.75$ and (b) $\epsilon_d=2$.
The dotted lines corresponds to 
$V=\epsilon_d\eta_c$, at which the engine reaches the Carnot efficiency.
The values of the remaining parameters are 
$\beta_L=1$, $\beta_R=0.4$, $\mu_L=V$, $\mu_R=0$, and $\Gamma=\Omega=0.005$, to enforce the weak
coupling (tight coupling) limit.
}
\label{fig:de}
\end{figure}


We now analyze the efficiency of the engine and the functionals 
$\sigma\Phi_C-2$ and $\sigma\Phi_E-2$ based on exact simulations.
To satisfy the resonant tunnelling condition, 
we fix $\Gamma=\Omega=0.005$, which results in 
$\Theta=0.72$. 
According to Eq. (\ref{eq:br_ed}), 
the necessary condition for TUR violation (while producing power) is $\beta_R\epsilon_d<0.544$.
To verify it, we further fix $\beta_R=0.4$ and vary $\epsilon_d$. 

In particular, we study two cases, $\epsilon_d=0.75$ and $\epsilon_d=4$,
corresponding to $\beta_R\epsilon_d<0.544$ and $\beta_R\epsilon_d>0.544$, respectively, with
results presented in Figs. \ref{fig:tur_te_small} and \ref{fig:tur_te}.
First, in panel (a) of both figures we illustrate the engine's efficiency. 
The weak coupling engine can reach the Carnot efficiency 
while the strong coupling engine shows a nonlinear behavior. 
At higher voltage, the engine ceases to operate as a thermoelectric generator. 
In Fig. \ref{fig:tur_te_small} (b), 
we indeed observe a violation of TUR in the resonant tunnelling regime, 
but the magnitude of violation is very small and the TUR is valid when the 
thermoelectric generator approaches its thermodynamic efficiency limit,
or operates in the strong coupling regime. 
In Fig. \ref{fig:tur_te} (b)
 we demonstrate that the TUR holds for both weak and strong coupling strengths, as $\epsilon_d$
is selected outside the appropriate range according to Eq. (\ref{eq:br_ed}). 

We also note from Fig. \ref{fig:tur_te} that the functionals 
$\sigma\Phi_C-2$ and $\sigma\Phi_E-2$ behave very similarly 
in the resonant tunnelling regime. This limit is sometimes referred to as the ``tight coupling regime",
with the currents and noises being proportional to each other,
 $\mathcal{T}_1^1 =\epsilon_d\mathcal{T}_1^0$,
 $\mathcal{T}_1^2\approx \epsilon_d^2\mathcal{T}_1^0$ and 
$\mathcal{T}_2^2\approx\epsilon_d^2\mathcal{T}_2^0$ 
according to Eq. (\ref{eq:t12}) in the limit of $\epsilon_d\gg\Omega, \Gamma$, 
thereby yielding $\Phi_C\approx\Phi_E$. 

It is also worthwhile to remark the range of voltage utilized in simulations. 
Since we let $\mu_L=V$ and $\mu_R=0$ in the analysis, 
the  efficiency $\eta$ of the thermoelectric engine is given by
\begin{equation}
\eta~=~V\frac{\langle j_C\rangle}{\langle j_E\rangle}.
\end{equation}
In the resonant tunnelling regime, 
due to $\langle j_E\rangle_{res}=\epsilon_d\langle j_C\rangle_{res}$, 
we get $\eta=V/\epsilon_d$ as confirmed by exact results for 
$\Omega=\Gamma=0.005$ depicted in Figs. \ref{fig:tur_te_small} (a) and \ref{fig:tur_te} (a). 
This imposes a constraint on the range of voltage $V$ that 
we can vary in the simulations since we should fulfill the requirement 
$\eta\leq\eta_c\equiv1-\beta_R/\beta_L$. 
At strong couplings, $\eta$ depicts a turn-over instead of a linearly increasing behavior. 
However, if we further increase the voltage, we find that $\eta$ becomes negative 
\cite{Bijay.15.PRB}, implying that the system no longer operates as an engine. 
Thus, one should be cautious when choosing the range of voltage $V$ in simulations,
ensuring that the system produces power.


In Fig. \ref{fig:con} we further depict a contour map of $\eta-\eta_c$ and $\sigma\Phi_E-2$
as a function of $V$ and $\epsilon_d$, 
crossing into the domain where the system no longer operates as an engine.
%
We focus on the weak coupling limit by recalling that (i) in this regime, $\sigma\Phi_E-2$ and $\sigma\Phi_C-2$ behave very similarly, and (ii) the simple form, $\eta=V/\epsilon_d$, holds such that the thermoelectric engine can reach the 
Carnot efficiency (marked as the black dashed line in Fig. \ref{fig:con} (a)) through adjusting the values of $V$ and $\epsilon_d$. 
When $\epsilon_d$ is large,  
we find that in the parameter regime where the system behaves as an engine, with $\eta\leq\eta_c$
 (region to the left of the dashed line in Fig. \ref{fig:con} (a)), 
the functional $\sigma\Phi_E-2$ is positive, implying that the TUR is valid. 

We note that it is merely impossible to distinguish in the contour map
the violations of TUR in the functional
regime of thermoelectric engines (to the left of the dashed line), which
 take place at small values of $\epsilon_d$, since the magnitude of violation is quite small.
%
In contrast, in panel (b) we do observe that $\sigma\Phi_E-2$ is negative over a broad range---yet
in fact in the regime where the system does not operate as an engine 
(to the right of the dashed line in panel(a)).

Although the engine studied here can attain the Carnot efficiency, 
it comes at the price of divergent current fluctuations, and consequently 
the relative uncertainty diverges. 
As a result, we observe spikes in Fig. \ref{fig:con} (b), 
precisely along the line $\eta=\eta_c$. 
(The reason why they form spikes instead of a continuous line is that we discretize $V$ and $\epsilon_d$ in obtaining the contour map) 

In Fig. \ref{fig:de} we clearly illustrate the divergent behavior of 
fluctuations and $\sigma\Phi_E-2$ as we approach the Carnot efficiency.
By fixing the value of $\epsilon_d$ and varying the voltage $V$, we find that $\sigma\Phi_E-2$  
diverges at $V=\epsilon_d\eta_c$, where the efficiency of the engine reaches the Carnot efficiency. 
An analogous behavior holds for $\sigma\Phi_C-2$.
In fact, using Eq. (\ref{eq:sigma}), we can rewrite the inequality $\sigma \Phi_{\alpha}\geq 2$ as
\bea
\frac{\langle\langle j_{\alpha}^2\rangle\rangle}{\langle j_{\alpha}\rangle } 
\geq \frac{2}{\beta_L} \frac{\eta}{\eta-\eta_c},
\label{eq:TURe}
\eea
with $\beta_L$ as the inverse temperature of the cold bath.
This inequality, derived first in Ref. \cite{Pietzonka.18.PRL} 
for Markovian systems, points that if one were to operate a continuous engine at finite
power close to the Carnot efficiency, then power fluctuations would diverge at least as 
$1/(\eta_c-\eta)$.
Our analysis and simulations show that this power-constancy-efficiency constraint
holds in our quantum system when approaching the Carnot limit, 
though it can be violated in regimes where the efficiency 
of thermoelectric engines is below the Carnot limit.
This intriguing fact calls for further explorations over the validity of
Eq. (\ref{eq:TURe}) in non-Markovian models.

We conclude that the TUR is satisfied for noninteracting thermoelectric generators 
when reaching the Carnot efficiency. Hence, no quantum effects can be utilized 
to circumvent Eq. (\ref{eq:TURe}) and tame power fluctuations near optimal efficiency.


\section{Summary}\label{sec:5}

We questioned whether the so-called thermodynamic uncertainty relation holds in noninteracting 
quantum thermoelectric junctions.
Invalidating the TUR potentially allows overcoming 
a bound on performance characteristics tying efficiency, power, and power fluctuations.
We identified the root of TUR violations, the range of parameters 
where it can take place in charge conducting and thermoelectric junctions, and the 
impact of the failure of the TUR on thermoelectric performance.

TUR violation stems from the existence of a nonvanishing ``quantum"  component of current noises, 
which results from correlated exchange of two electrons. 
Considering resonant tunnelling junctions, we proved that the TUR can be violated in both charge conducting junctions and thermoelectric generators within a certain range of parameters.
We illustrated our findings using the serial double quantum dot system. 
Exact numerical results confirmed our theoretical predictions. 
In particular, we showed that in systems with multiple thermodynamic affinities, 
the TUR can be violated---but only when the system is operated away from the optimal efficiency limit, 
or outside the functional regime.

Our analytical results such as Eq. (\ref{eq:tur_ineq}), which identifies a window for
 TUR violations in thermoelectric junctions, seem particular or cumbersome, 
yet they are effective for a significant class of problems. 
Our derivation assumed (i) quantum coherent transport obeying Eq.  (\ref{eq:CGF}), (ii) resonant tunnelling transport, i.e. weak coupling of the system to the reservoirs, 
and (iii) a degenerate orbital energy. Assumption (i) can be justified in the low temperature regime, 
(ii) and (iii) are frequently adopted in the analysis of quantum dots or molecular junctions 
thermoelectricity. Our predictions can therefore be tested within present technology. 

Taking into account electron-electron or electron-phonon interactions in quantum machines 
may invalidate our findings, received while considering noninteracting thermoelectric engines. Future work will be focused on  many-body steady-state quantum machines, 
and the search for systems or regimes of operation where one can use quantum effects 
to suppress current fluctuations without compromising the efficiency and output power 
\cite{Holubec.18.PRL}.
Deriving a fundamental quantum bound on performance replacing the classical relation,
Eq. (\ref{eq:tur}) remains an intriguing challenge.

\begin{acknowledgments}
J. Liu and D. Segal acknowledge support from  
the Natural Sciences and Engineering Research Council (NSERC) of Canada Discovery Grant
and the Canada Research Chairs Program.
\end{acknowledgments}


\renewcommand{\theequation}{A\arabic{equation}}
\setcounter{equation}{0}  
\section*{Appendix: ~Solution of (\ref{eq:tur_ineq}) for TUR violations in thermoelectric junctions}

Expanding the hyperbolic functions involved in Eq. (\ref{eq:tur_func_2}), we find
\begin{eqnarray}
\coth\frac{Y}{2} &\simeq& \frac{2}{Y}+\frac{Y}{6}-\frac{Y^3}{360},\nonumber\\
\tanh\frac{\beta_R\epsilon_d}{2} &\simeq& \frac{\beta_R\epsilon_d}{2}-\frac{\beta_R^3\epsilon_d^3}{24},\nonumber\\
\tanh\frac{\beta_R\epsilon_d-Y}{2} &\simeq& \frac{\beta_R\epsilon_d-Y}{2}-\frac{(\beta_R\epsilon_d-Y)^3}{24}.
\end{eqnarray}
In deriving the above expansions, we have already set $\mu_R=0$ and let $\beta_L(\epsilon_d-\mu_L)=\beta_R\epsilon_d-Y$. Inserting the above equations into Eq. (\ref{eq:tur_func_2}) we get
\begin{eqnarray}
\sigma\Phi_C-2 &\simeq& Y^2\Big[\frac{1}{6}-\frac{Y^2}{360}+\frac{\Theta}{2}\Big(-\frac{1}{2}+\frac{1}{8}\beta_R^2\epsilon_d^2\nonumber\\
&&-\frac{1}{8}\beta_R\epsilon_dY+\frac{Y^2}{24}\Big)\Big].
\end{eqnarray}
Recall, $\Theta\equiv\mathcal{T}_2^0/\mathcal{T}_1^0$ is a property of the junction.
The inequality $\sigma\Phi_C-2<0$ then leads to Eq.  (\ref{eq:tur_ineq}) after some rearrangements.
Let us organize Eq.  (\ref{eq:tur_ineq}), which identifies TUR violation, as
\bea
aY^2-bY+c < 0 
\label{eq:A1}
\eea
with 
\bea
a&=&\left(15\Theta-2\right)
\nonumber\\
b&=& 45\beta_R\epsilon_d\Theta
\nonumber\\
c&=&-(180-45\beta_R^2\epsilon_d^2)\Theta+120.
\label{eq:A2}
\eea
This is a parabola and the inequality can be interrogated by studying its roots $\lambda_{1,2}$,
\bea
\lambda_{1,2}=\frac{(45 \beta_R\epsilon_d\Theta   \mp \sqrt\Delta)} {2(15\Theta-2)}.
\label{eq:A3}
\eea
Here, $\Delta=b^2-4ac$.
Since for strong enough thermodynamical forces dissipation is excessive and the TUR 
should be satisfied, we conclude that $a>0$ and that Eq. (\ref{eq:A1}) corresponds to 
\begin{equation}
\frac{45\beta_R\epsilon_d\Theta-\sqrt{\Delta}}{2(15\Theta-2)}~<~Y~<~\frac{45\beta_R\epsilon_d\Theta+\sqrt{\Delta}}{2(15\Theta-2)}.
\label{eq:A4}
\end{equation}
%
In order to satisfy Eq. (\ref{eq:A4}) we should require that 
\begin{eqnarray}\label{eq:Delta_d}
\Delta &=& 2025\beta_R^2\epsilon_d^2\Theta^2-4\left(120-180\Theta+45\beta_R^2\epsilon_d^2\Theta\right)\nonumber\\
&&\times\left(15\Theta-2\right)\nonumber\\
&=& 15\Big[(720-45\beta_R^2\epsilon_d^2)\Theta^2-\left(576-24\beta_R^2\epsilon_d^2\right)\Theta\nonumber\\
&&+64\Big]>0.
\end{eqnarray}
%
As the ratio $\Theta$ is independent of temperature, this inequality for $\Delta$ should be regarded as a constraint on possible values of $\beta_R\epsilon_d$. It gives
\begin{equation}
\beta_R^2\epsilon_d^2~<~\frac{576\Theta-720\Theta^2-64}{24\Theta-45\Theta^2},
\label{eq:A8}
\end{equation}
together with the conditions
\begin{equation}
\left.
\begin{array}{c}
24\Theta-45\Theta^2~<~0 \\
576\Theta-720\Theta^2-64~<~0
\end{array}\right\}
~\Rightarrow~\Theta>2/3
\end{equation}
as $\beta_R^2\epsilon_d^2$ is non-negative.

Eqs. (\ref{eq:A8}) together with $\Theta>2/3$ are necessary  
conditions for TUR violations in thermoelectric junctions. 
While interesting by itself, we are looking here for TUR violations {\it under the restriction that
the thermoelectric junction produces power}.
To operate the system as a thermoelectric engine we require that
$Y<0$, see Eq. (\ref{eq:cc}).  
We therefore additionally demand that $\lambda_1<0$, 
or equivalently that $\Delta>(45\beta_R\epsilon_d\Theta)^2$ 
such that we can identify a validity range for $Y$ as $\lambda_1<Y<0$ from Eq. (\ref{eq:A4}). 

\begin{figure}[tbp!]
  \includegraphics[width=0.85\columnwidth]{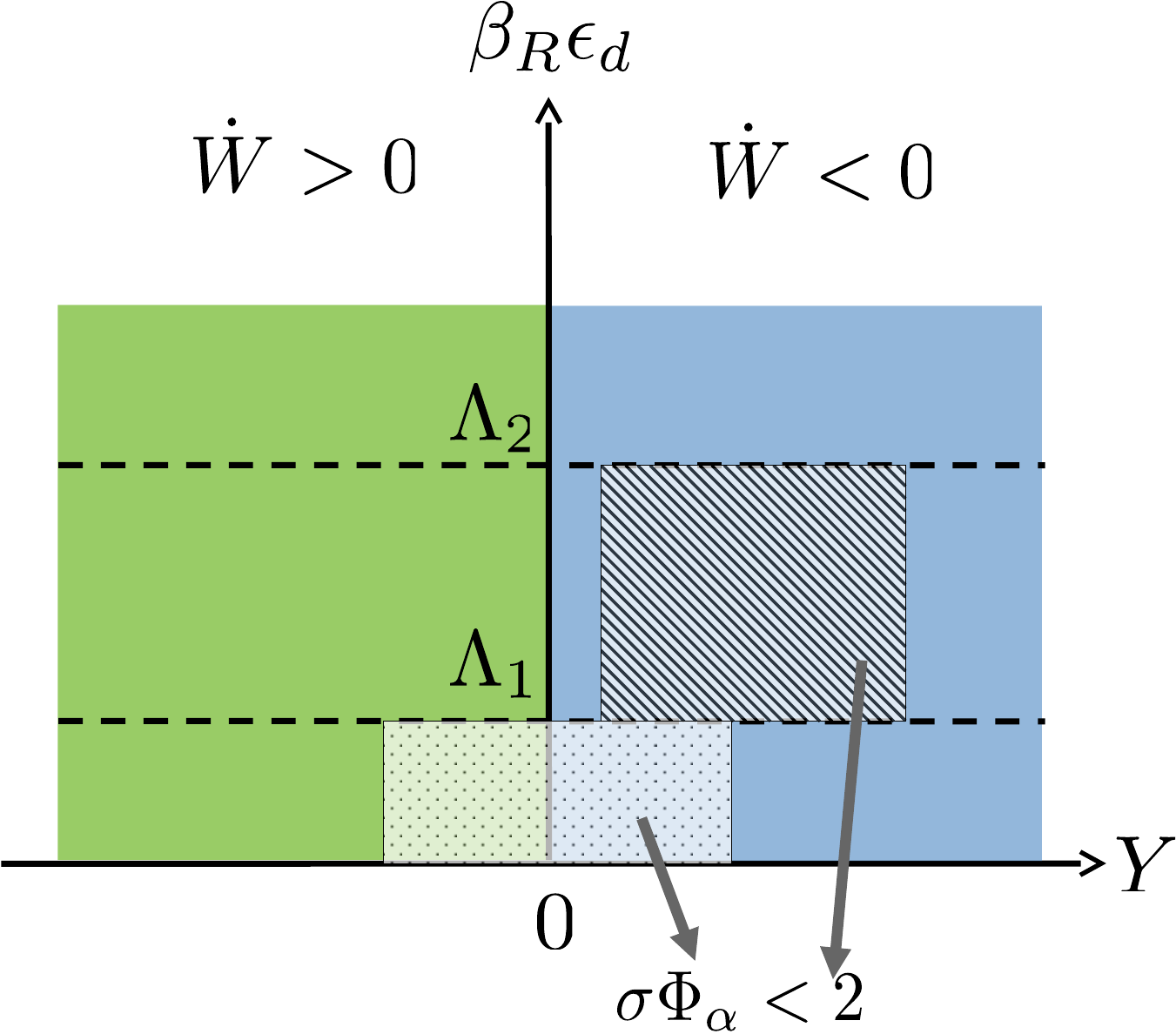}
\caption{ Illustration of condition $\lambda_1<Y<\lambda_2$ as well as  (\ref{eq:A10}) and (\ref{eq:A11}),
where the TUR is violated, using dotted and diagonally patterned boxes, respectively.
Power is generated in the $Y<0$ domain.}
\label{fig:box}
\end{figure}

The condition $\lambda_1<0$ is fulfilled when $120-180\Theta+45\beta_R^2\epsilon_d^2\Theta<0$, 
as can be seen from Eq. (\ref{eq:Delta_d}).
Even more simply, since  in Eq. (\ref{eq:A1}) 
$a$ and $b$ are positive, and we limit $Y$ to the negative domain, the inequality can be
only satisfied if $c<0$,
\begin{equation}
0~<~\beta_R^2\epsilon_d^2~<~\frac{12\Theta-8}{3\Theta}=\Lambda_1^2,
\label{eq:A10}
\end{equation}
where we have used the fact that $12\Theta-8>0$.

Eq. (\ref{eq:A10}) is a necessary condition to overcoming the TUR for a thermoelectric engine 
producing power. This situation is depicted in Fig. \ref{fig:box} with $\lambda_1<Y<\lambda_2$ and 
$\beta_R\epsilon_d<\Lambda_1$ defining a window for TUR violation (dotted patterned box).
 
Since $\frac{12\Theta-8}{3\Theta}<\frac{576\Theta-720\Theta^2-64}{24\Theta-45\Theta^2}$
within the validity range of the ratio $\Theta>2/3$, Eq. (\ref{eq:A10}) is  more restrictive
than Eq. (\ref{eq:A8}), as expected. 
In contrast to Eq. (\ref{eq:A10}), the TUR can be violated but the junction does not 
produce power when 
\begin{equation}
\frac{12\Theta-8}{3\Theta}~\le~\beta_R^2\epsilon_d^2~<~\frac{576\Theta-720\Theta^2-64}{24\Theta-45\Theta^2}=\Lambda_2^2,
\label{eq:A11}
\end{equation}
since now $\lambda_1>0$ and no negative $Y$ can satisfy Eq.  (\ref{eq:tur_ineq}). This situation
is illustrated in Fig. \ref{fig:box} within the diagonally patterned window.

 
 

%

\end{document}